\documentclass[12pt]{iopart}
\usepackage{multirow}
\usepackage{setstack}
\usepackage{csquotes}
\usepackage{amsfonts}
\usepackage{cases}
\usepackage{graphicx}
\usepackage{xcolor}
\usepackage[colorlinks=true,linkcolor=blue,citecolor=red]{hyperref}%
\usepackage{cite}
\usepackage{url}

\newcommand\restr[2]{{% we make the whole thing an ordinary symbol
  \left.\kern-\nulldelimiterspace % automatically resize the bar with \right
  #1 % the function
  \right|_{#2} % this is the delimiter
  }}

\def\R{{\mathbb R}}
\def\C{{\mathbb C}}
\def\B{{\mathcal B}}
\def\P{{\mathcal P}}
\def\T{{\mathcal T}}
\def\r{\mathbf{r}}

\def\V{{\mathsf V}}

\newcommand{\eqref}[1]{(\ref{#1})}

\usepackage{iopams}  

\begin{document}

\title{Spectral branch points of the Bloch-Torrey operator}

\author{Nicolas Moutal}

\address{Laboratoire de Physique de la Mati{\`e}re Condens{\'e}e, CNRS -- Ecole Polytechnique, IP Paris,
91128 Palaiseau, France}
\ead{nicolas.moutal@gmail.com}

\author{Denis S. Grebenkov}
\address{Laboratoire de Physique de la Mati{\`e}re Condens{\'e}e, CNRS -- Ecole Polytechnique, IP Paris,
91128 Palaiseau, France}
\ead{denis.grebenkov@polytechnique.edu}

\begin{abstract}
We investigate the peculiar feature of non-Hermitian operators,
namely, the existence of spectral branch points (also known as
exceptional or level crossing points), at which two (or many)
eigenmodes collapse onto a single eigenmode and thus loose their
completeness.  Such branch points are generic and produce
non-analyticities in the spectrum of the operator, which, in turn,
result in a finite convergence radius of perturbative expansions based
on eigenvalues and eigenmodes that can be relevant even for Hermitian
operators.  We start with a pedagogic introduction to this phenomenon
by considering the case of $2\times 2$ matrices and explaining how the
analysis of more general differential operators can be reduced to this
setting.  We propose an efficient numerical algorithm to find spectral
branch points in the complex plane.  This algorithm is then employed
to show the emergence of spectral branch points in the spectrum of the
Bloch-Torrey operator $-\nabla^2 - igx$, which governs the time
evolution of the nuclear magnetization under diffusion and precession.
We discuss their mathematical properties and physical implications for
diffusion nuclear magnetic resonance experiments in general bounded
domains.
\end{abstract}

\submitto{\JPA}

\section{Introduction}

Hermitian (or self-adjoint) operators play the central role in quantum
and classical physics by governing a broad variety of natural
phenomena, such as diffusion, wave propagation, or evolution of a
quantum state.  Spectral properties of these operators have been
thoroughly investigated
\cite{Reed,Kato,Baumgartel,Helffer2013a,Zinn-Justin1989a,Hall}.  In many
practically relevant cases, a Hermitian operator has a discrete (or
pure point) spectrum of real eigenvalues, while the associated
eigenfunctions (also called eigenmodes or eigenvectors) form a
complete orthogonal basis in the underlying functional space.  As an
expansion of a function on that basis reduces the operator to a
multiplication operator, the related spectral expansions are commonly
used, e.g., in quantum mechanics and in stochastic theory.  Even
though Hermitian operators form an ``exceptional'' set of operators
(in the same sense as real numbers are ``exceptional'' among complex
numbers), they are most often encountered in applications.  In turn,
more general non-Hermitian operators that may possess a discrete,
continuous or even empty spectrum, present a richer variety of
spectral properties \cite{Moiseyev2011a}.  Even if the spectrum is
discrete, some eigenvalues may coalesce at specific branch points
(also known as exceptional points or level crossing points), at which
the completeness of eigenmodes is generally lost, thus failing
conventional spectral expansions \cite{Kato,Helffer2013a}.  As
spectral branch points lie in the complex plane, this failure can be
unnoticed but still relevant even for Hermitian operators.  For
instance, Bender and Wu studied an anharmonic (quartic) quantum
oscillator governed by the Hamiltonian $\B(g) = - d^2/dx^2 + x^2 + g
x^4$ and discovered an infinity of branch points that cause divergence
of the perturbation series in powers of $g$ for the ground-energy
state \cite{Bender1969a} (see 
\cite{Simon1970a,Delabaere1997a,Shapiro19,Shapiro22} for further
details and extensions).  Even if the parameter $g$ is real and the
operator $\B(g)$ is thus Hermitian, the accumulation of branch points
near the origin prohibits a perturbative approach to this problem.  In
other words, the existence of branch points may crucially affect the
spectral expansions of Hermitian operators as well.

Peculiar properties of spectral branch points have been studied for
various non-Hermitian operators (see, e.g.,
\cite{Berry2004a,Heiss2004a,Seyranian05,Kirillov05,Rubinstein2007a,Cartarius2007a,Cejnar2007a,Klaiman2008a,Chang2009a,Ceci2011a,Shapiro17,Grosfjeld19}
and references therein).  
While a branch point often involves two eigenvalues, higher-order
branch points offer even richer mathematical structure and physical
properties.  For instance, Jin considered two symmetrically coupled
asymmetric dimers and discussed phase transitions when encircling the
exceptional points \cite{Jin18}.  Hybrid exceptional points that
exhibit different dispersion relations along different directions in
the parameter space, were studied on a two-state system of coupled
ferromagnetic waveguides with a bias magnetic field \cite{Zhang18},
while high-order exceptional points in supersymmetric arrays of
coupled resonators or waveguides under a gradient gain and loss were
discussed in \cite{Zhang20} (see also
\cite{Demange12,Melanathuru22}).

Apart from being an object of intensive mathematical investigations,
spectral branch points have recently found numerous physical
implications.  For instance, Xu {\it et al.} reported an experimental
realization of the transfer of energy between two vibrational modes of
a cryogenic optomechanical device that arises from the presence of an
exceptional point in the spectrum of that device \cite{Xu16}.
More generally, a dynamical encircling of an exceptional point in
parameter space may lead to the accumulation of a geometric phase and
thus realize a robust and asymmetric switch between different modes,
as shown experimentally for scattering through a two-mode waveguide
with suitably designed boundaries and losses \cite{Doppler16} (see
also \cite{Zhang18b,Feilhauer20}).  Yoon {\it et al.}  constructed two
coupled silicon-channel waveguides with photonic modes that transmit
through time-asymmetric loops around an exceptional point in the
optical domain, taking a step towards broadband on-chip optical
devices based on non-Hermitian topological dynamics \cite{Yoon18}.
Along the same line, Schumer {\it et al.}  designed a structure that
allows the lasing mode to encircle a non-Hermitian exceptional point
so that the resulting state transfer reflects the unique topology of
the associated Riemann surfaces associated with this singularity
\cite{Schumer22}.  This approach aimed to provide a route to
developing versatile mode-selective active devices and to shed light
on the interesting topological features of exceptional points.  Hassan
{\it et al.} theoretically analyzed the behavior of two coupled states
whose dynamics are governed by a non-Hermitian Hamiltonian undergoing
cyclic variations in the diagonal terms \cite{Hassan17}.  They
obtained analytical solutions that explain the asymmetric conversion
into a preferred mode and the chiral nature of this mechanism.
However, the latest experiments on a fibre-based photonic emulator
showed that chiral state transfer can be realized without encircling
an exceptional point and seems to be mostly attributed to the
non-trivial landscape of the energy Riemann surfaces \cite{Nasari22}.

In this paper, we investigate the spectral branch points of the
Bloch-Torrey differential operator $\mathcal{B}(g) = -\nabla^2 - ig
x$, acting on a subspace of square-integrable functions on a bounded
Euclidean domain $\Omega\subset \R^d$ \cite{Torrey1956a}.  For $g\in
\R$, this non-Hermitian operator describes diffusion and precession of
nuclear spins under a magnetic field gradient
\cite{Callaghan1991a,Price2009a,Jones2011a,Axelrod2001a,Grebenkov2007a,Grebenkov2009a,Grebenkov2016b,Kiselev2017a}.
While spectral properties of the Bloch-Torrey operator have been
studied in the past
\cite{Stoller1991a,Swiet1994a,Huerlimann1995a,Grebenkov2014b,Grebenkov2017a,Herberthson2017a,Grebenkov2018b,Almog2018a,Almog2019a,Moutal2020a,Grebenkov2021a},
a systematic analysis of its branch points is still missing.  In fact,
we are aware of a single work by Stoller {\it et al.}
\cite{Stoller1991a}, who discussed spectral branch points of the
Bloch-Torrey operator on an interval with reflecting endpoints.  In
addition, few numerical examples of branch points for some planar
domains were given in \cite{Moutal2020a}.  We start with a pedagogic
introduction to spectral branch points in Sec. \ref{sec:general} by
considering a basic example of $2\times 2$ matrices (this introductory
section may be skipped by an expert reader).  Section
\ref{section:BT_bifurcation} describes our main results.  We first
look at the spectral branch points of the Bloch-Torrey operator
$\mathcal{B}(g)$ in parity-symmetric domains.  In order to deal with
arbitrary domains, we present a numerical algorithm for determining
spectral branch points in the complex plane.  In particular, we
illustrate that branch points generally lie outside the real axis when
the parity symmetry is lost.  In Sec. \ref{sec:discussion}, we outline
how the presence of branch points implies a finite convergence radius
of perturbative expansions.  In particular, we estimate this radius
for the Bloch-Torrey operator in two domains and discuss its immediate
consequences for diffusion nuclear magnetic resonance (NMR)
experiments.  Section \ref{sec:conclusion} concludes the paper with a
summary of main results and future perspectives.

\section{Pedagogic introduction to spectral branch points}
\label{sec:general}

In this section, we provide a pedagogic introduction to spectral
branch points from a general point of view for non-expert readers.
This is a simplified sketch of classical descriptions that can be
found in most textbooks on spectral theory
\cite{Kato,Baumgartel,Helffer2013a,Moiseyev2011a,Seyranian}.

Let $\mathcal{B}(g)$ denote a linear operator on a vector space $E$
that depends analytically on a complex parameter $g$.  We assume that
$\mathcal{B}(g)$ has a discrete spectrum for any $g\in\mathbb{C}$, and
$\{\lambda_n(g)\}$ denotes the set of its eigenvalues, i.e., the poles
of the resolvent $(\mathcal{B}(g) - \lambda \mathcal{I})^{-1}$.
Intuitively, one can expect that each $\lambda_n(g)$ depends
``smoothly'' on $g$, except for some points, at which two (or many)
eigenvalues coincide.  For a Hermitian operator, crossing of
eigenvalues does not alter the spectral properties and does keep
analytical dependence on $g$.  In contrast, crossing of two
eigenvalues of a non-Hermitian operator at some point $g_0$ results in
a non-analytical dependence on $g$ in a vicinity of $g_0$.  Moreover,
the behavior of the corresponding eigenmodes drastically changes at
$g_0$, as described below.

\subsection{Spectral branch points as complex branch points of a multi-valued function}

The spectrum $\lbrace \lambda_n(g) \rbrace$ of the operator $\mathcal
B(g)$ is often obtained by solving an equation of the form
\begin{equation}  \label{eq:F}
F(\lambda,g)=0, 
\end{equation}
where $F$ is an analytical function of $\lambda$ and $g$.  For a
matrix, this function is simply its characteristic polynomial, while a
differential operator generally yields a transcendental function $F$.
For example, the operator $\mathcal B = -\partial_x^2 + g$ on the
interval $(0,1)$ with Dirichlet boundary condition leads to the
spectral equation: $F(\lambda,g) = \sin(\sqrt{\lambda-g})=0$.

\begin{figure}[pt]
\centering
\includegraphics[width=0.99\linewidth]{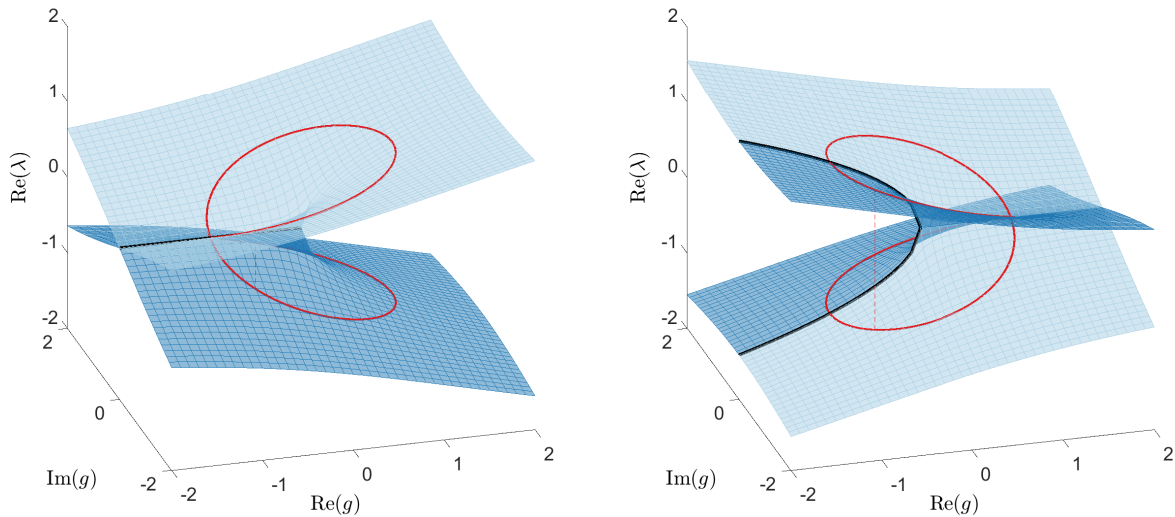} % branchement_racine_dg.eps}
\caption{
Plot of the real part (left) and imaginary part (right) of the
multi-valued function $\lambda(g)=\pm \sqrt{g}$ (``$+$'' sheet in
light blue and ``$-$'' sheet in dark blue).  The cut along the real
negative semi-axis is represented as a thick black line.  Each sheet
is discontinuous at the cut but both of them form a single continuous
surface $\lambda(g)$.  The red contour depicts a full $4\pi$ turn
around $g=0$ on the surface $\lambda(g)$.  After a $2\pi$ turn from
the point $(\lambda_0,g_0)$, one reaches the point $(\lambda_0,-g_0)$,
as indicated by the red dashed line.  }
\label{fig:branchement_racine}
\end{figure}

To illustrate how branch points may result from the spectral equation
\eqref{eq:F}, let us consider a simple example:
\begin{equation}  
F(\lambda,g) = \lambda^2 - g = 0\;.
\end{equation}
This equation can be inverted to obtain $\lambda$ as a function of $g$
but the inversion of the square function makes $\lambda(g)$ a
multi-valued function with two possible values (i.e., two sheets in
the complex plane):
\begin{equation}  \label{eq:lambda12}
\lambda_1(g) = \sqrt{g}\;, \qquad \lambda_2(g)= - \sqrt{g}\; .\\
\end{equation}
The multi-valued character of $\lambda(g)$ is closely related to the
absence of the unique determination of the argument of a complex
number and the need of a cut in the complex plane.  In what follows,
we employ the usual convention that the cut is along the negative real
semi-axis.  In other words, the square root is defined as follows:
\begin{equation}
\sqrt{\rho \,e^{i\theta}} = \sqrt\rho \, e^{i\theta/2}\;, \qquad \rho > 0\,, \;-\pi < \theta \leq \pi\;.
\end{equation}
This choice makes the real part of $\sqrt g$ a continuous function
when $g$ crosses the cut (i.e., when $\theta$ jumps from $\pi$ to
$-\pi$).  However, the imaginary part of $\sqrt g$ is not continuous
and jumps from $i\sqrt\rho$ to $-i\sqrt \rho$.  Figure
\ref{fig:branchement_racine} shows the multi-valued function
$\lambda(g) =\pm \sqrt g$.  One can see two sheets are individually
discontinuous at the cut, but both sheets taken together form a
continuous surface.  By performing a $2\pi$ turn around $g=0$, one
goes from one sheet to the other, and a full $4\pi$ turn is required
to go back to the initial point.

\begin{figure}[pt]
\centering
\includegraphics[width=0.99\linewidth]{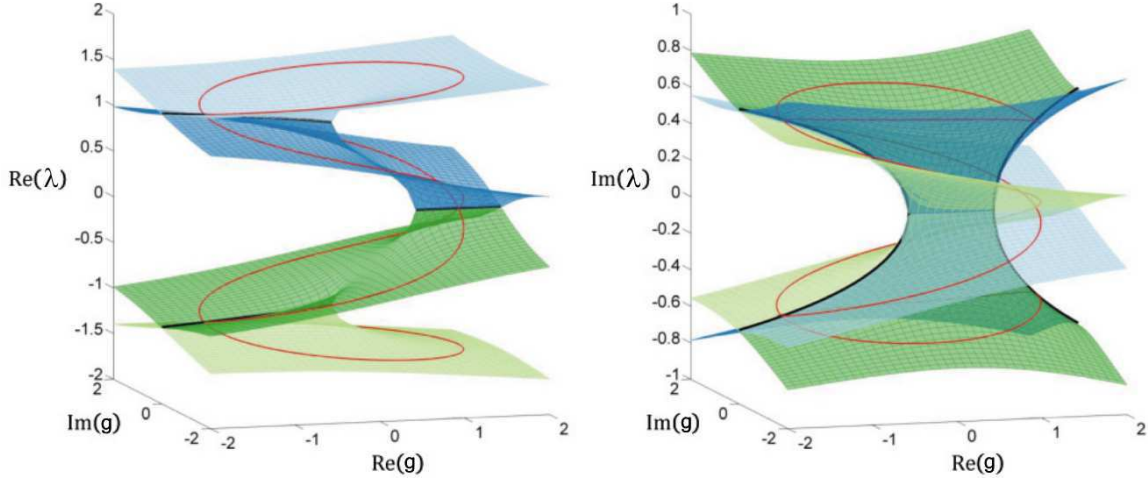} % branchement_complexe_dg.pdf}
\caption{
Plot of the real part (left) and imaginary part (right) of the
multi-valued function $\lambda(g)=\pm \sqrt{\pm\sqrt g +1}$.  Four
sheets are shown by different colors: $++$ in light blue, $+-$ in dark
blue, $-+$ in light green and $--$ in dark green.  The red contour
depicts a full $6\pi$ turn around all branch points, and illustrates
that all sheets are connected to each other to form a unique
multi-valued surface.  }
\label{fig:branchement_complexe}
\end{figure}

This basic example can be extended to other multi-valued functions.
For instance, the inversion of the equation
\begin{equation}  
F(\lambda,g)=(\lambda^2-1)^2 - g = 0
\end{equation}
makes $\lambda(g)$ a multi-valued function with four possible values
(i.e., four sheets in the complex plane):
\begin{subequations}  \label{eq:lambda14}
\begin{eqnarray}
&\lambda_1(g)=\sqrt{\sqrt{g}+1}\;, \qquad \lambda_2(g)=\sqrt{-\sqrt{g}+1}\;,\\
&\lambda_3(g)=-\sqrt{\sqrt{g}+1}\;,\qquad \lambda_4(g)=-\sqrt{-\sqrt{g}+1}\;.
\end{eqnarray}
\end{subequations}
This multi-valued function, shown on
Fig. \ref{fig:branchement_complexe}, exhibits three branch points, at
which two sheets coincide: $\lambda_1(0)=\lambda_2(0)=1$,
$\lambda_3(0)=\lambda_4(0)=-1$, and $\lambda_2(1)=\lambda_4(1)=0$
(note that the first two branch points occur at the same value $g =
0$).  Although it is visually more complicated than the square root
function, it is essentially the combination of three $\sqrt g$-type
branch points that connect four sheets together.  By circling around
all branch points (a contour shown in red), one goes through all
sheets and reaches the initial point after a $6\pi$ turn.

In general, spectral branch points of an operator $\mathcal{B}(g)$ are
related to branch points of a complex multi-valued function.  From a
mathematical point of view, the eigenvalues $ \lambda_1(g),
\lambda_2(g),\ldots$ can be interpreted as different sheets of a
unique multi-valued function $\lambda(g)$ that results from the
inversion of the transcendental eigenvalue equation \eqref{eq:F}.  In
the next subsection, we explain by computations with $2\times 2$
matrices that two crossing eigenvalues behave as $\sqrt{g-g_0}$ in a
vicinity of their branch point.

\subsection{Matrix model}
\label{section:bifurcation_matrix}

\begin{figure}[t]
\centering
\includegraphics[width=0.99\linewidth]{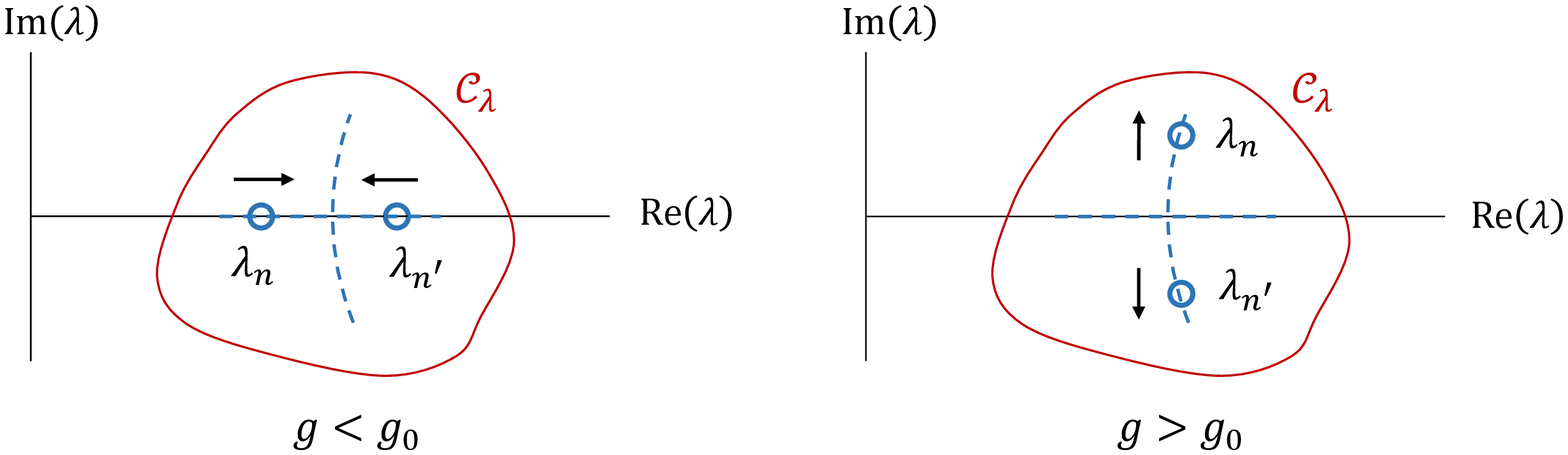} % projector_bifurcation_dg-crop.pdf}
\caption{
Illustration of the transition from two real eigenvalues (left) to a
complex conjugate pair (right).  At $g=g_0$, $\lambda_n$ and
$\lambda_{n'}$ coincide in the complex plane.  By integrating the
resolvent $(\mathcal{B}(g) - \lambda \mathcal{I})^{-1}$ over a contour
$\mathcal C_\lambda$, one obtains a two-dimensional projector $\Pi(g)$
that is analytical in $g$ because the resolvent is analytical outside
of poles (i.e., away from the eigenvalues of $\mathcal{B}(g)$).}
\label{fig:bifurcation}
\end{figure}

We illustrate the phenomenon of spectral branch points on the simplest
case of a $2\times 2$ matrix (see also, e.g.,
\cite{Heiss2004a,Gunther2007a}).  Although an operator $\B(g)$ acting on an
infinite-dimensional vector space cannot be reduced to a matrix, the
coalescence of two eigenmodes and eigenvalues is essentially captured
by a computation on a vector space of dimension $2$ (note that the
coalescence of a larger number of eigenmodes, that can be realized
with larger matrices, leads to branch points of higher order and
involves a higher-dimensional space, see
\cite{Jin18,Zhang18,Zhang20} and references therein; this case is not
considered here).  To explain this point, we follow the suggestion by
B.~Helffer illustrated on Fig. \ref{fig:bifurcation}.  Let us choose
an integration contour $\mathcal{C}_\lambda$ in the complex
$\lambda$-plane that circles around two simple (non-degenerate)
eigenvalues $\lambda_n$ and $\lambda_{n'}$.  We assume that these (and
only these) eigenvalues coalesce at $g = g_0$.  Since the spectrum is
discrete, it is possible to choose the contour $\mathcal{C}_\lambda$
such that no other eigenvalue cross it over $|g - g_0| < \epsilon$ for
a given small enough $\epsilon > 0$.  By integrating the resolvent
$(\mathcal{B}(g) - \lambda \mathcal{I})^{-1}$ of the operator
$\mathcal{B}(g)$ over the contour $\mathcal{C}_\lambda$, one obtains a
two-dimensional projector $\Pi(g)$ over the space spanned by the
associated eigenmodes $v_n$ and $v_{n'}$, at least for $g \neq g_0$.
Note that $\Pi(g)$ is a function of the parameter $g$ with values in
the infinite-dimensional space of continuous operators over the vector
space $E$.  Such a resolvent expansion in the vicinity of a branch
point is well known since the 1960s.  For the illustrative purposes of
this section, we focus on the main ideas and skip mathematical
details, subtleties, and proofs that can be found in classical
textbooks on operator theory (see, e.g.,
\cite{Baumgartel,Helffer2013a}).  

Since the integration contour $\mathcal{C}_\lambda$ does not cross any
eigenvalue, the resolvent is an analytical function of $\lambda$ and
$g$ over $\mathcal{C}_\lambda$, therefore $\Pi(g)$ is an analytical
function of $g$.  In particular, the image of $\Pi(g)$ is
two-dimensional, even at the branch point $g=g_0$.  As one will see,
this does not imply that there are still two linearly independent
eigenmodes $v_n$ and $v_{n'}$ at that point.  The restriction of the
operator $\mathcal{B}(g)$ to the image of $\Pi(g)$ yields a $2\times
2$ matrix $\mathsf A(g)$.  If there is no other spectral branch point
over the considered range of $g$, the restriction of the operator to
the kernel of $\Pi(g)$ is analytical, therefore the non-analyticity of
the spectrum of $\mathcal{B}(g)$ is fully captured by the matrix
$\mathsf A(g)$ as claimed above.

We consider a $2\times 2$ matrix of the general form
\begin{equation}
\mathsf A(g)=\left[\begin{array}{cc}\lambda_0 + a& b\\c &\lambda_0 -a \end{array}\right]\;,
\end{equation}
where $\lambda_0,a,b,c$ are smooth functions of $g$ (the smoothness
results from the analyticity of the projector $\Pi(g)$).  One can
easily compute its eigenvalues $\lambda_{\pm}$ and eigenvectors
$\mathsf X_{\pm}$:
\begin{eqnarray}
&\lambda_{\pm}=\lambda_0 \pm\sqrt{d}\;,\qquad \mathsf X_{\pm}=\left[\begin{array}{c} b\\ \pm\sqrt{d}-a \end{array}\right]\;,
\end{eqnarray}
where $d = bc + a^2$.  If $d \ne 0$, the eigenvalues are distinct
while the associated eigenvectors are linearly independent.  In turn,
if $d(g_0)=0$ at some point $g_0$, the eigenvalues coincide at
$g=g_0$.  We first consider a Hermitian matrix and then we show how
the general non-Hermitian case differs qualitatively.

(i) If $\mathsf A(g)$ is Hermitian for all values of $g$, then
$a\in\mathbb{R}$ and $c=b^*$, so that $d=|b|^2+a^2$ is real and
non-negative.  Furthermore, the condition $d(g_0) = 0$ implies
$a(g_0)=b(g_0)=c(g_0)=0$.  This also yields $d'(g_0)=0$, while
$d''(g_0)\neq0$ in general, so that the eigenvalues in a vicinity of
$g_0$ are approximately equal to
\begin{equation} 
\lambda_\pm \approx \lambda_0(g_0) \pm \sqrt{d''(g_0)/2}\, (g-g_0)\;.
\end{equation}
Thus one can draw two main conclusions: (i) the spectrum does not
present non-analytical branch points, the eigenvalues merely cross
each other at $g=g_0$; (ii) the dimension of the eigenspace
$E_{\lambda_0}$ at $g=g_0$ is $2$.  Such a coalescence point is
called {\it diabolic point}
\cite{Berry84,Seyranian05,Kirillov05,Gunther06}.

(ii) Now we turn to the non-Hermitian case.  The function $d(g)$ takes
complex values and crosses $0$ at $g=g_0$ with a non-zero derivative
$d'(g_0)$.  The phases of $\lambda_{\pm}-\lambda_0$ undergo a $\pi/2$
jump when $g$ passes through the critical value $g_0$ and the absolute
value of $\lambda_{\pm}-\lambda_0$ behaves typically as
$\sqrt{d'(g_0)(g-g_0)}$ for $g$ close to $g_0$.  In particular, if
$d(g)$ is real (positive for $g<g_0$ and negative for $g>g_0$), and if
$\lambda_0(g_0)$ is real, one obtains in a vicinity of the critical
value $g_0$:
\begin{subequations}
\begin{eqnarray}
g<g_0 \qquad & \left\lbrace
\begin{array}{l}
\mathrm{Re}(\lambda_{\pm})\approx \lambda_0(g_0) \pm \sqrt{d'(g_0)(g_0-g)} \\ \mathrm{Im}(\lambda_{\pm})=0
\end{array}\right.  \vspace{6pt}\\
g>g_0 \qquad 
&\left\lbrace\begin{array}{l}
\mathrm{Re}(\lambda_{\pm})= \lambda_0(g_0) \\ \mathrm{Im}(\lambda_{\pm})\approx\pm \sqrt{d'(g_0)(g-g_0)}
\end{array}\right.
\end{eqnarray}
\end{subequations}

At the critical value $g=g_0$, the matrix $\mathsf A(g_0)$ is in
general not diagonalizable.  Without loss of generality, let us assume
that $b(g_0)\neq 0$.  The matrix $\mathsf A(g_0)$ can then be reduced
to a Jordan block with an eigenvector $\mathsf X_0$ and a generalized
eigenvector $\mathsf Y_0$:
\begin{subequations}
\begin{eqnarray}
&\mathsf A(g_0)\mathsf X_0= \lambda_0 \mathsf X_0 \;, \qquad \mathsf X_0=\left[\begin{array}{c} b(g_0)\\-a(g_0) \end{array}\right]\;, \\
&\mathsf A(g_0)\mathsf Y_0= \lambda_0 \mathsf Y_0 + \mathsf X_0 \;, \qquad \mathsf Y_0= \left[\begin{array}{c} 0\\1\end{array}\right]\;.
\end{eqnarray}
\end{subequations}
Note that since the derivative of $\sqrt{d(g)}$ is infinite at
$g=g_0$, one has
\begin{equation}
\mathsf Y_0=\restr{\frac{d \mathsf X_\pm}{d \lambda_\pm}}{g=g_0}\;,
\end{equation}
where the derivative yields the same result for $(\mathsf
X_+,\lambda_+)$ and $(\mathsf X_-,\lambda_-)$.  Moreover, if $\mathsf
A(g)$ is a symmetric matrix (i.e. $b=c$), then $\mathsf X_0^T \mathsf
X_0 = a^2(g_0) + b^2(g_0) = d(g_0) = 0$, i.e., $\mathsf X_0$ is
``orthogonal'' to itself for the real scalar product.

In comparison to the Hermitian case, there are two main conclusions:
(i) the spectrum is non-analytical at $g=g_0$; (ii) the eigenvectors
$\mathsf X_\pm$ of $\mathsf A(g)$ collapse onto a {\it single}
eigenvector $\mathsf X_0$ as $g\to g_0$, and the matrix $\mathsf
A(g_0)$ can be reduced to a Jordan block with a generalized
eigenvector $\mathsf Y_0$ given by the rate of change of the
eigenvectors $\mathsf X_\pm$ with their corresponding eigenvalues
$\lambda_\pm$, evaluated at the critical point $g=g_0$.  Such a
coalescence point is called {\it exceptional point}.

\begin{figure}[tp]
\centering
\includegraphics[width=0.99\linewidth]{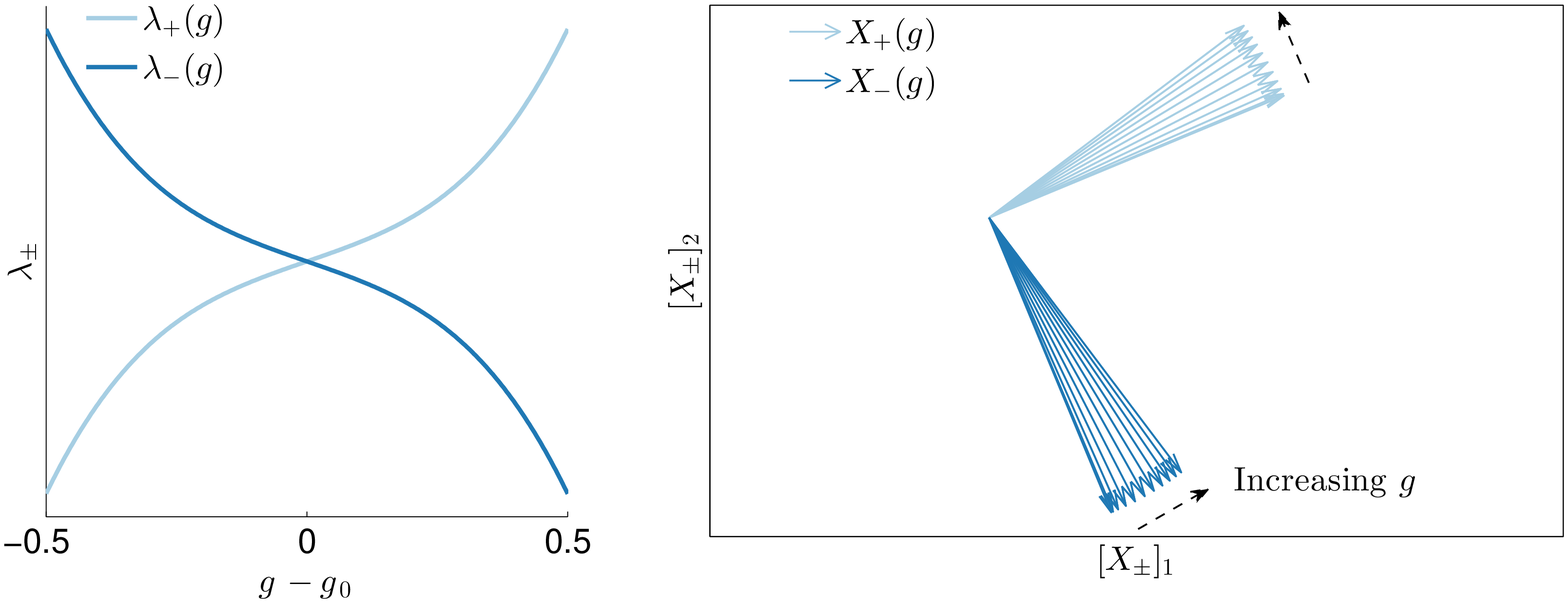} % bifurcation_2_2_hermitian_dg.eps}
\noindent\rule[5pt]{\linewidth}{0.4pt}
\includegraphics[width=0.99\linewidth]{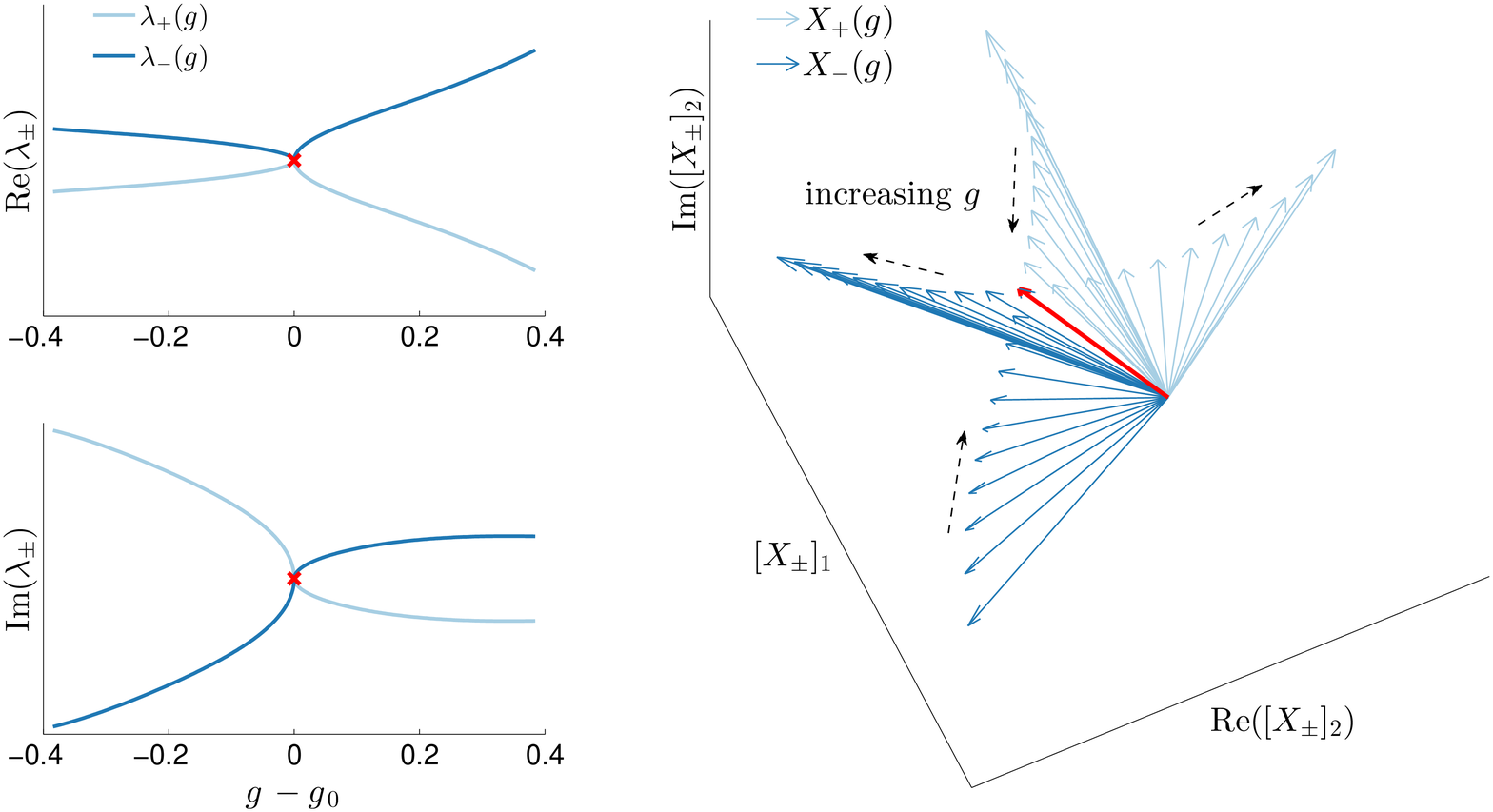} % bifurcation_2_2_non_hermitian_dg.eps}
\caption{
Eigenvalues and eigenvectors of a general $2\times 2$ matrix $\mathsf
A(g)$.  (Top) Hermitian case: the eigenvalues $\lambda_\pm(g)$ cross
each other at $g=g_0$, the eigenvectors $X_\pm(g)$ are always
orthogonal to each other and do not exhibit any particular behavior at
$g=g_0$.  (Bottom) non-Hermitian case: the eigenvalues
$\lambda_\pm(g)$ exhibit a typical square-root behavior near $g=g_0$
(indicated by a red cross), the eigenvectors $X_\pm(g)$ collapse onto
a single vector at $g = g_0$ (indicated by a red arrow).  Note that
$g$ was sampled with additional points near $g_0$ for a better
visualization of the branching.  The complex $X_\pm$ vectors were
plotted with the convention that $\mathrm{arg}([X_\pm]_1)=0$. }
\label{fig:bifurcation_2_2}
\end{figure}

We summarize the results for the Hermitian and non-Hermitian cases
graphically on Fig. \ref{fig:bifurcation_2_2}.  We emphasize that the
dichotomy ``Hermitian $\leftrightarrow$ no branching'' versus
``non-Hermitian $\leftrightarrow$ branching'' is specific to
two-dimensional matrices.  For instance, if one considers a $4\times4$
matrix made of two $2\times2$ blocks where one is Hermitian and the
other is non-Hermitian, then the eigenvalues of the Hermitian block do
not exhibit any branch point when they cross, even if the whole matrix
is not Hermitian.  This somewhat artificial example shows that there
is no general relation between branch points and the non-Hermitian
property, except that the spectrum of a Hermitian operator never
branches.  By reducing the full operator to a low-dimensionality
matrix on the subspace associated to the crossing point, one can make
precise statements about branching and Hermitian character, as we did
in this two-dimensional case.  The ``translation'' of the above
conclusions to the case of the Bloch-Torrey operator and their
consequences on spectral decompositions are detailed in
Sec. \ref{section:app_bifurcation_spectral_decomposition}.

\section{The Bloch-Torrey operator}
\label{section:BT_bifurcation}

In this section we apply the previous considerations to the
Bloch-Torrey operator which governs the time evolution of the
transverse magnetization $m(t,\mathbf r)$ in diffusion NMR experiments
\cite{Callaghan1991a,Price2009a,Jones2011a,Axelrod2001a,Grebenkov2007a,Grebenkov2009a,Grebenkov2016b,Kiselev2017a}.
For a given Euclidean domain $\Omega \subset \R^d$ ($d \geq 1$) with a
smooth enough boundary $\partial \Omega$, the transverse magnetization
of the spin-bearing particles obeys the Bloch-Torrey equation
\cite{Torrey1956a}
\begin{equation}
\partial_t \, m(t,\mathbf r) = D \nabla^2 m(t,\mathbf r) + i\gamma G x \, m(t,\mathbf r),
\end{equation}
subject to some initial magnetization $m(0,\mathbf r)$ and to Neumann
boundary condition:
\begin{equation}  \label{eq:BC}
\restr{\mathbf n \cdot \nabla m}{\partial \Omega} = 0\;,
\end{equation}
where $\mathbf n$ is the inward normal vector at the boundary, $D$ is
the diffusion coefficient, $\gamma$ is the gyromagnetic ratio, and $G$
is the amplitude of the magnetic field gradient applied along some
spatial direction $x$ (here the gradient is considered to be
time-independent).  The Laplace operator $\nabla^2$ governs diffusion
of the spin-bearing particles while the imaginary ``potential''
$-i\gamma Gx$ describes the magnetization precession in the transverse
plane due to the applied magnetic field gradient.  Denoting $g =
\gamma G/D$, the solution of this equation can be formally written as
$m(t,\mathbf r) = e^{-\mathcal B(g) Dt} m(0,\mathbf r)$, where we
introduced the Bloch-Torrey operator
\begin{equation}
\mathcal B(g) = -\nabla^2  - igx  
\label{eq:BT_operator}
\end{equation}
that acts on functions on the domain $\Omega$, subject to the boundary
condition (\ref{eq:BC}) (for a rigorous mathematical definition, see
e.g. \cite{Grebenkov2018b,Almog2018a,Almog2019a}).
Even though the parameter $g$ is real in the context of diffusion NMR,
we consider the general case $g\in \C$.

In the one-dimensional setting, $\mathcal B(g)$ is more commonly known
as the (complex) Airy operator \cite{Helffer2013a,Grebenkov2017a}.
For a half-line $\Omega = \mathbb{R}_+$, its eigenmodes are known
explicitly in terms of the Airy function, whereas the eigenvalues are
expressed in terms of its zeros
\cite{Stoller1991a,Grebenkov2014b,Grebenkov2017a} (see also
\cite{Voros1999a,Voros1999b} for relations to higher-order polynomial
potentials).  This is the only known example, for which the spectrum
of the Bloch-Torrey operator is fully explicit.  Even for an interval,
$\Omega = (a,b)$, one has to solve numerically a transcendental
equation to determine the eigenvalues (see
\cite{Stoller1991a,Grebenkov2014b} for details).
Despite this difficulty, Stoller {\it et al.}  managed to discover and
accurately characterize branch points in the spectrum
\cite{Stoller1991a}.

In higher dimensions ($d \geq 2$), a linear combination of the
Laplace operator $-\nabla^2$ and a coordinate-linear potential term
$-igx$ along a single spatial direction naturally suggests a spatial
splitting of the setup into this (longitudinal) direction and its
orthogonal (transversal) coordinate hyperplane.  One can therefore
represent the Bloch-Torrey operator as a sum of the corresponding
one-dimensional Airy-type operator and the Laplace operator in the
transversal hyperplane.  The mode coupling (and mixing) occurs through
the boundary conditions (\ref{eq:BC}).
The spectral properties of the Bloch-Torrey operator have been studied
in different geometric settings, including bounded, unbounded and
periodic domains
\cite{Swiet1994a,Huerlimann1995a,Grebenkov2014b,Grebenkov2017a,Herberthson2017a,Grebenkov2018b,Almog2018a,Almog2019a,Moutal2020a,Grebenkov2021a}.
For instance, the asymptotic behavior of the eigenvalues and
eigenfunctions in the semi-classical limit $g\to \infty$ has been
analyzed.  In turn, the existence of branch points in the spectrum of
the Bloch-Torrey operator in Euclidean domains with $d \geq 2$ remains
unknown, except for a numerical example given in \cite{Moutal2020a}.

Throughout this section we assume that the domain $\Omega$ is bounded.
Under this assumption, the gradient term is a bounded perturbation of
the Laplace operator that ensures the existence of a discrete set of
eigenpairs (eigenmode and eigenvalue) $(v_n,\lambda_n)$, indexed by
$n=1,2,\ldots$ and ordered according to the growing real part of
$\lambda_n$.  If $g$ is not a branch point (see below), the eigenmodes
are orthonormal, $(v_n|v_k)=\delta_{n,k}$, with respect to the
bilinear form
\begin{equation}  \label{eq:scalar}
(f|g)=\int_\Omega f(\mathbf r) g(\mathbf r) \,d\mathbf r\;.
\end{equation}
We recall that there is no complex conjugate in the definition of
$(\cdot | \cdot)$ because the Bloch-Torrey operator is not Hermitian.
This normalization implies that each eigenmode is defined up to a sign
(in contrast to Hermitian operators, there is no arbitrary phase
factor $e^{i\alpha}$).

We first show how spectral branch points appear for real values of $g$
in parity symmetric domains.  Interestingly, this branching phenomenon
is associated to a dramatic change in the behavior of the eigenmodes
of the Bloch-Torrey operator, with an abrupt transition from
delocalized to localized states (see below).  Then we turn to general
domains and present a simple numerical algorithm to detect branch
points in the complex $g$-plane.

\subsection{Parity-symmetric domains}

\begin{figure}[t]
\centering
\includegraphics[width=0.3\linewidth]{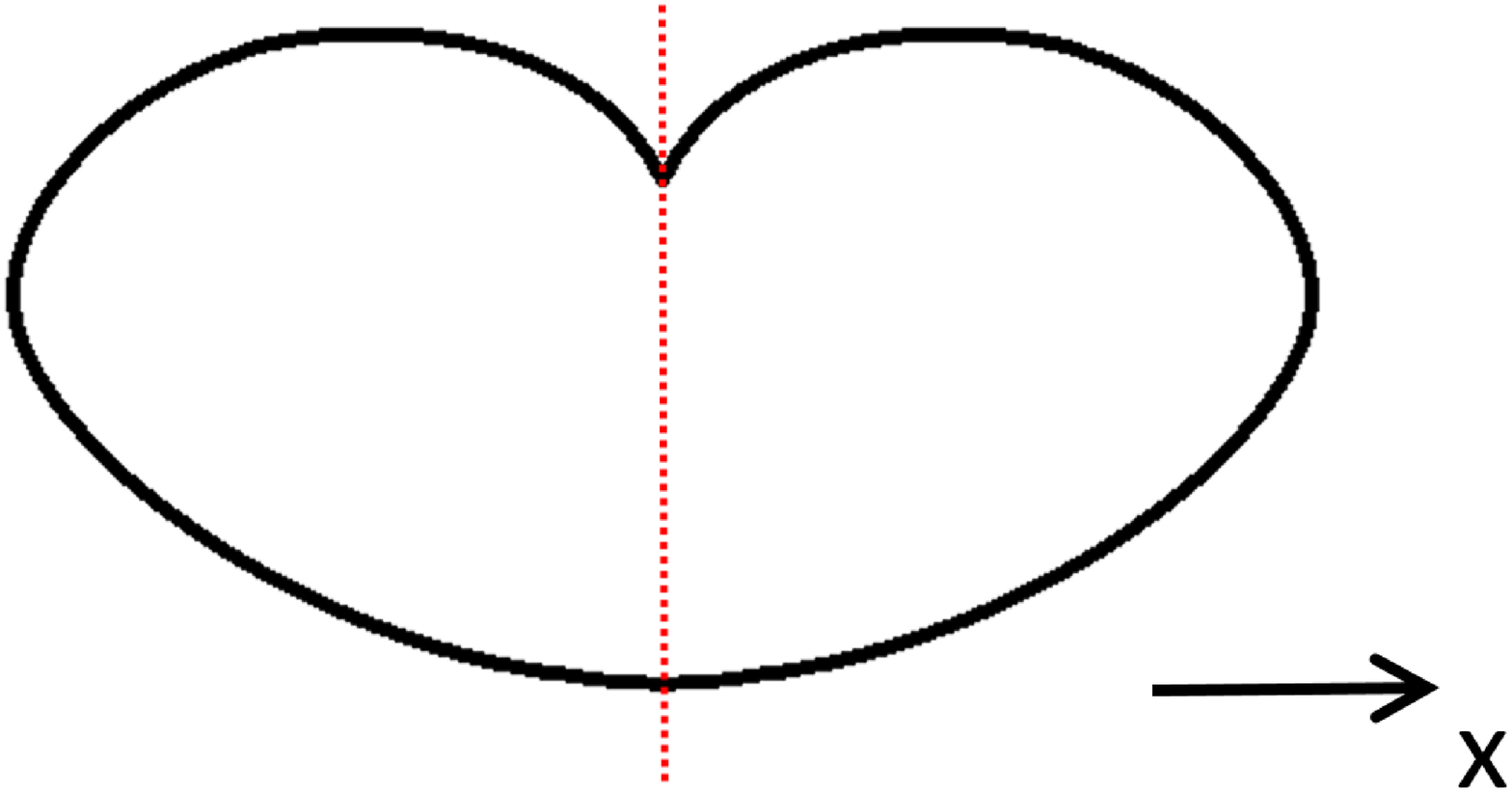} % domain1.eps}
\includegraphics[width=0.3\linewidth]{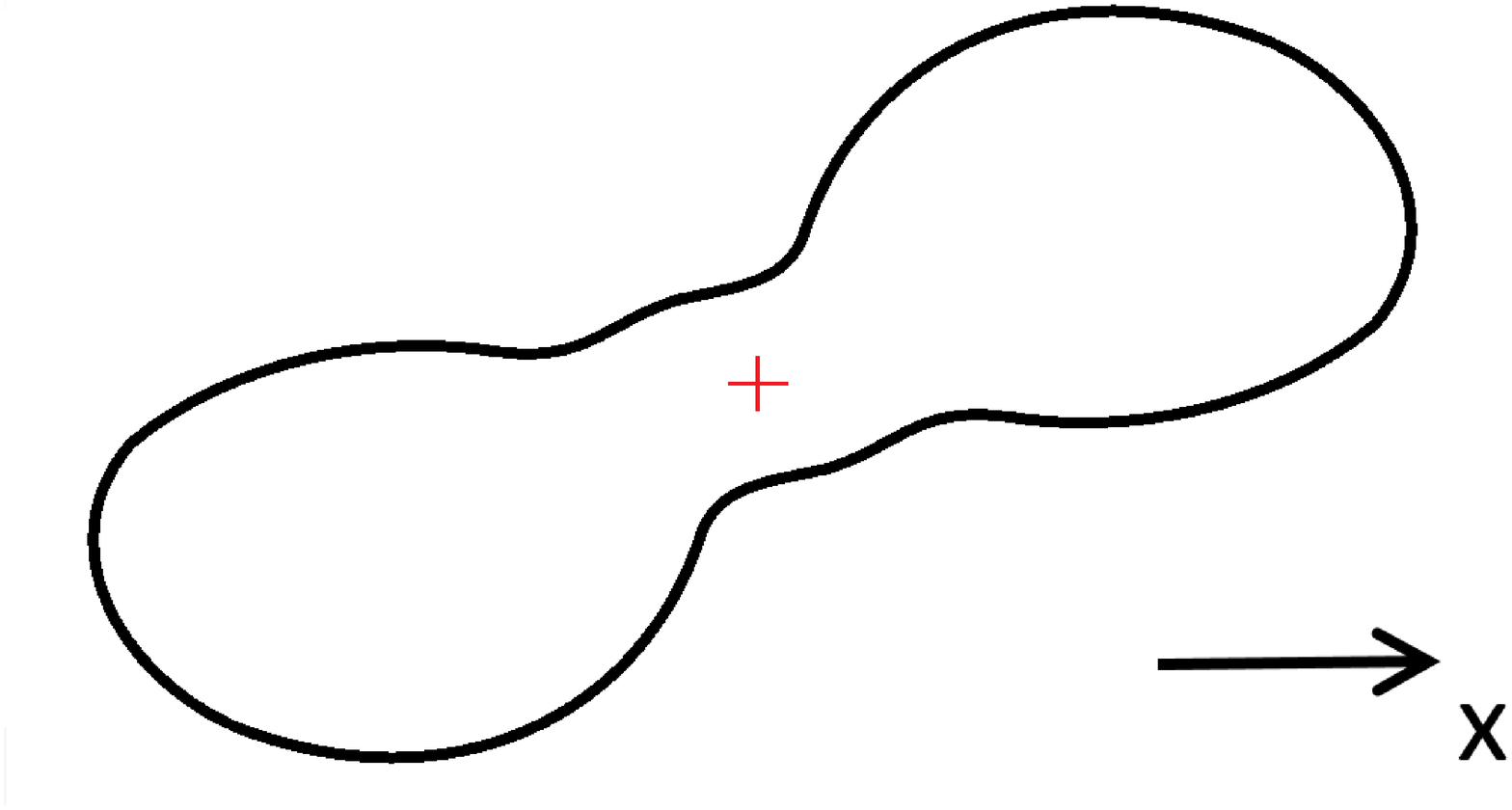} % domain2.eps}
\caption{
Illustration of two domains that are symmetric under $x$-parity:
(left) the domain is symmetric under mirror symmetry indicated by red
dotted line; (right) the domain is symmetric under central symmetry
indicated by red cross.}
\label{fig:symmetric_domains}
\end{figure}

Let us assume that the domain $\Omega$ is invariant by an isometric
transformation that reverses the $x$-axis (i.e., $\pi$ rotation around
an axis orthogonal to $\mathbf{e_x}$ or mirror symmetry with respect
to the plane orthogonal to $\mathbf{e_x}$), see
Fig. \ref{fig:symmetric_domains}.  We call this transformation
$x$-parity in short and denote it by $\mathcal{P}_x$.  Under the
combination $x$-parity and complex conjugation, the Bloch-Torrey
operator becomes
\begin{equation}
\mathcal{P}_x \mathcal{B}(g)^*=(-\nabla^2 - i g (-x))^* = \mathcal{B}(g^*)\;.
\end{equation}
In this subsection, we focus on the case of real $g$ so that the
Bloch-Torrey operator is invariant under $x$-parity plus complex
conjugation (since the complex conjugation is associted with time
reversal in quantum mechanics, one usually speaks about $\P\T$
symmetry).  Therefore, if $v_n$ is an eigenmode with an eigenvalue
$\lambda_n$, then $\mathcal{P}_x v_n^*$ is an eigenmode with the
eigenvalue $\lambda_n^*$.  This leads to two possible situations.

\begin{enumerate}
\item 
the eigenvalue $\lambda_n$ is real (and simple), so that $v_n = \pm
\mathcal{P}_x v_n^*$.  In this case, the eigenmode $v_n$ is
``symmetric'' in the sense that $|v_n|$ is invariant by $x$-parity.
Note that this is consistent with the previous paragraph: the
imaginary part of $\lambda_n$ is zero and the eigenmode is centered
around $x=0$.  We call such an eigenmode ``delocalized'' as it spreads
over the whole domain.

\item 
two eigenvalues $\lambda_n$ and $\lambda_{n'}$ form a complex
conjugate pair, so that $v_{n'}= \pm \mathcal{P}_x v_n^*$.  This means
that the eigenmode $v_{n'}$ is an $x$-parity ``reflection'' of $v_n$.
We call such an eigenmode ``localized'' as it is mainly concentrated
on one side of the domain and is almost zero on the other side.
\end{enumerate}

The transition from case (i) to case (ii), i.e. from two real
eigenvalues $\lambda_n$ and $\lambda_{n'}$ to a complex conjugate pair
may occur only if $\lambda_n$ and $\lambda_{n'}$ coincide at some
value $g_0$.  As illustrated on Fig. \ref{fig:bifurcation}, the
coalescence of two eigenvalues results in a branch point in the
spectrum, i.e., a non-analyticity.  Such branch points mark the
transition from delocalized eigenmodes (i) to localized eigenmodes
(ii) that are at the origin of the localization regime.  This
phenomenon was first discovered by Stoller \textit{et al.} for the
Bloch-Torrey operator in an interval with Neumann boundary condition
\cite{Stoller1991a}.

\begin{figure}[tp]
\centering
\includegraphics[width=0.49\linewidth]{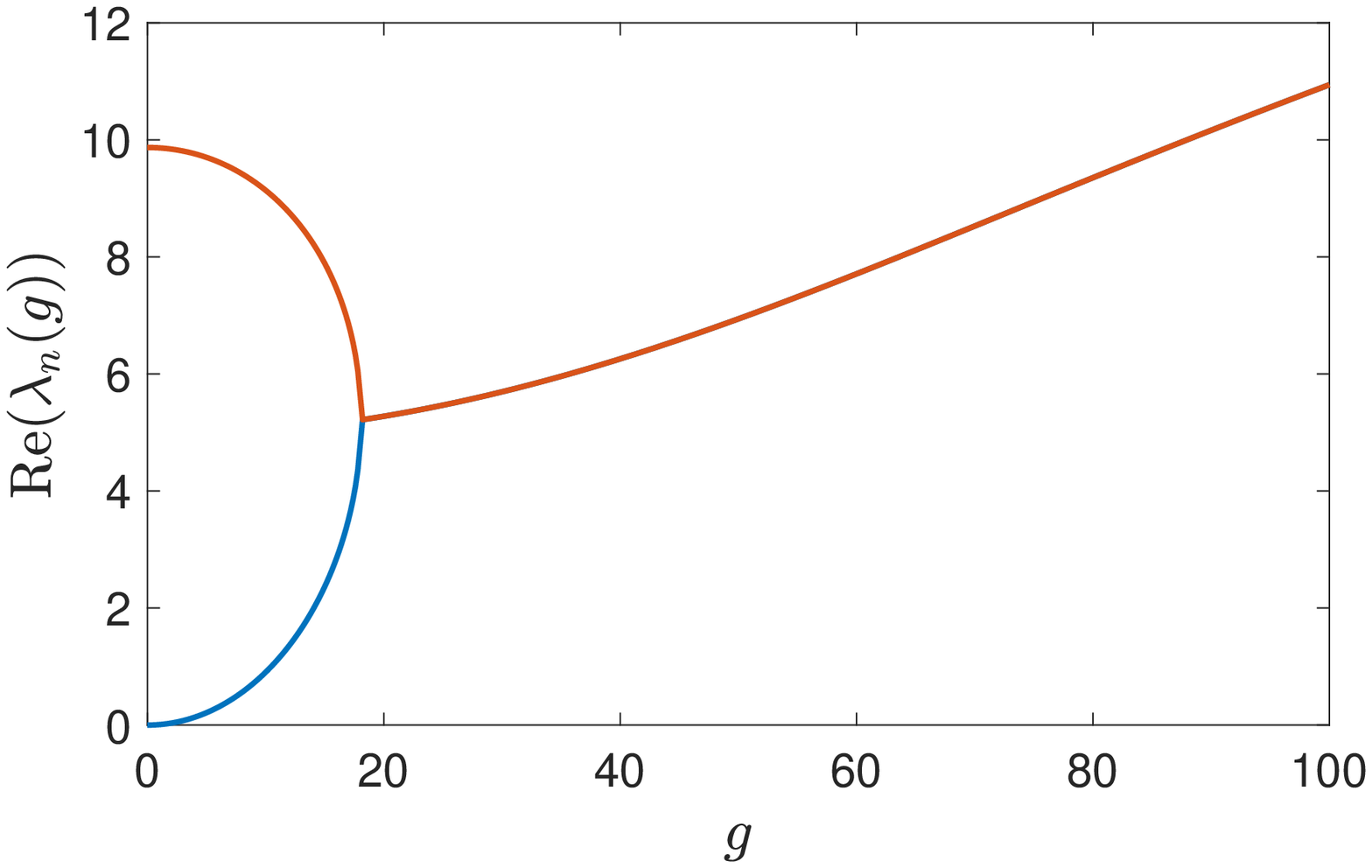} % 1d_lambda_real2.eps}
\includegraphics[width=0.49\linewidth]{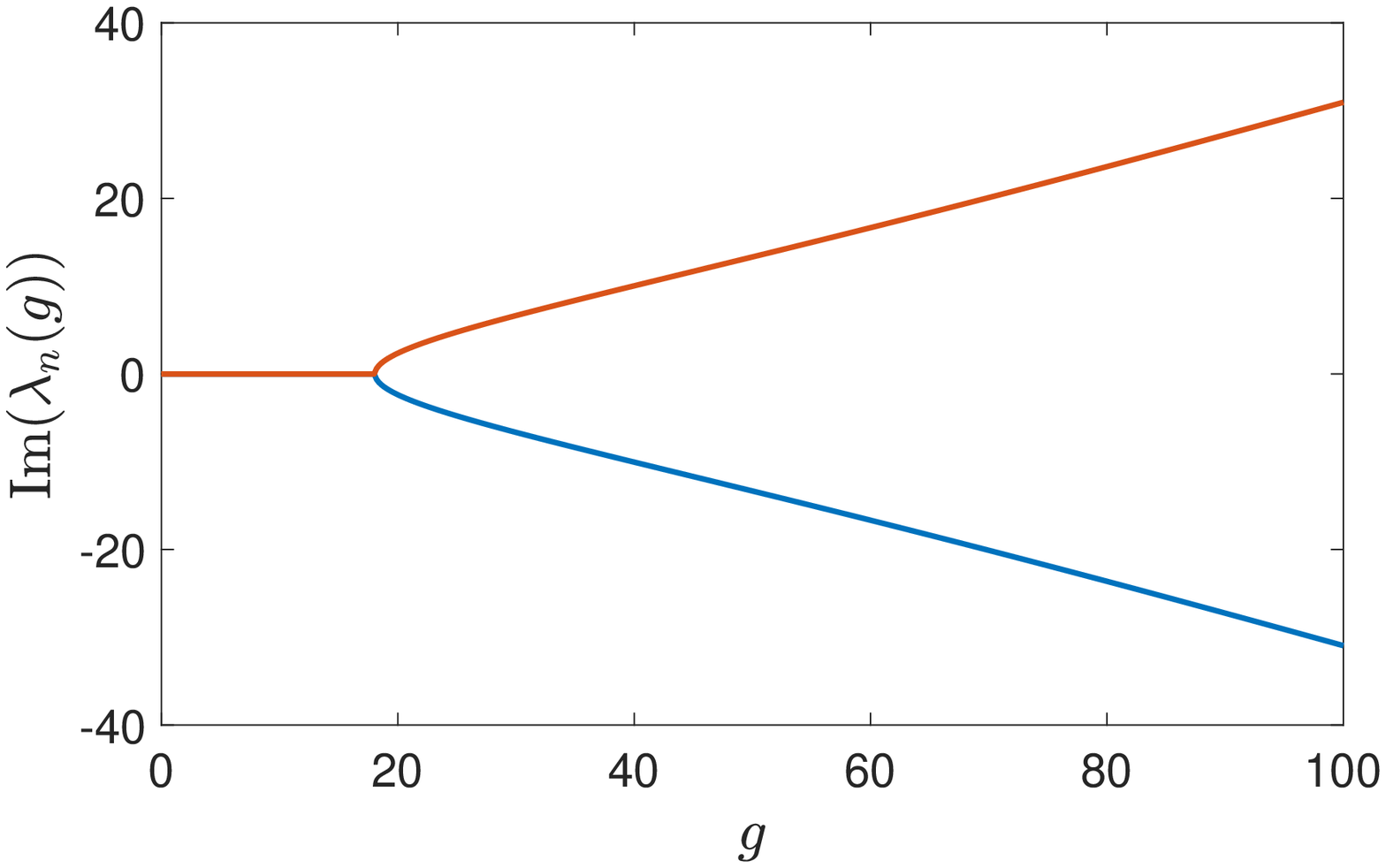} % {1d_lambda_imag2.eps}
\includegraphics[width=0.49\linewidth]{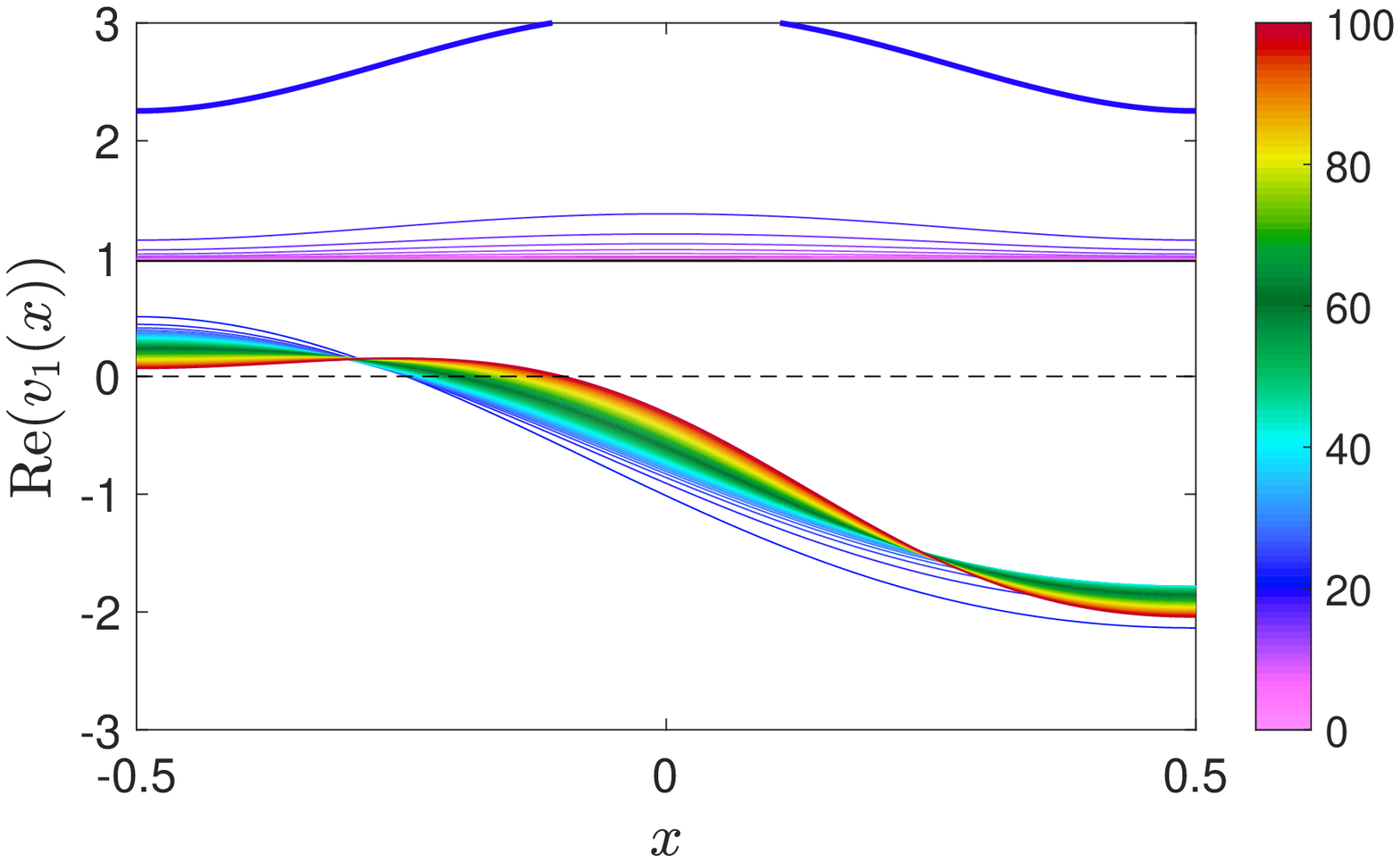} % {1d_v1_Re.eps}
\includegraphics[width=0.49\linewidth]{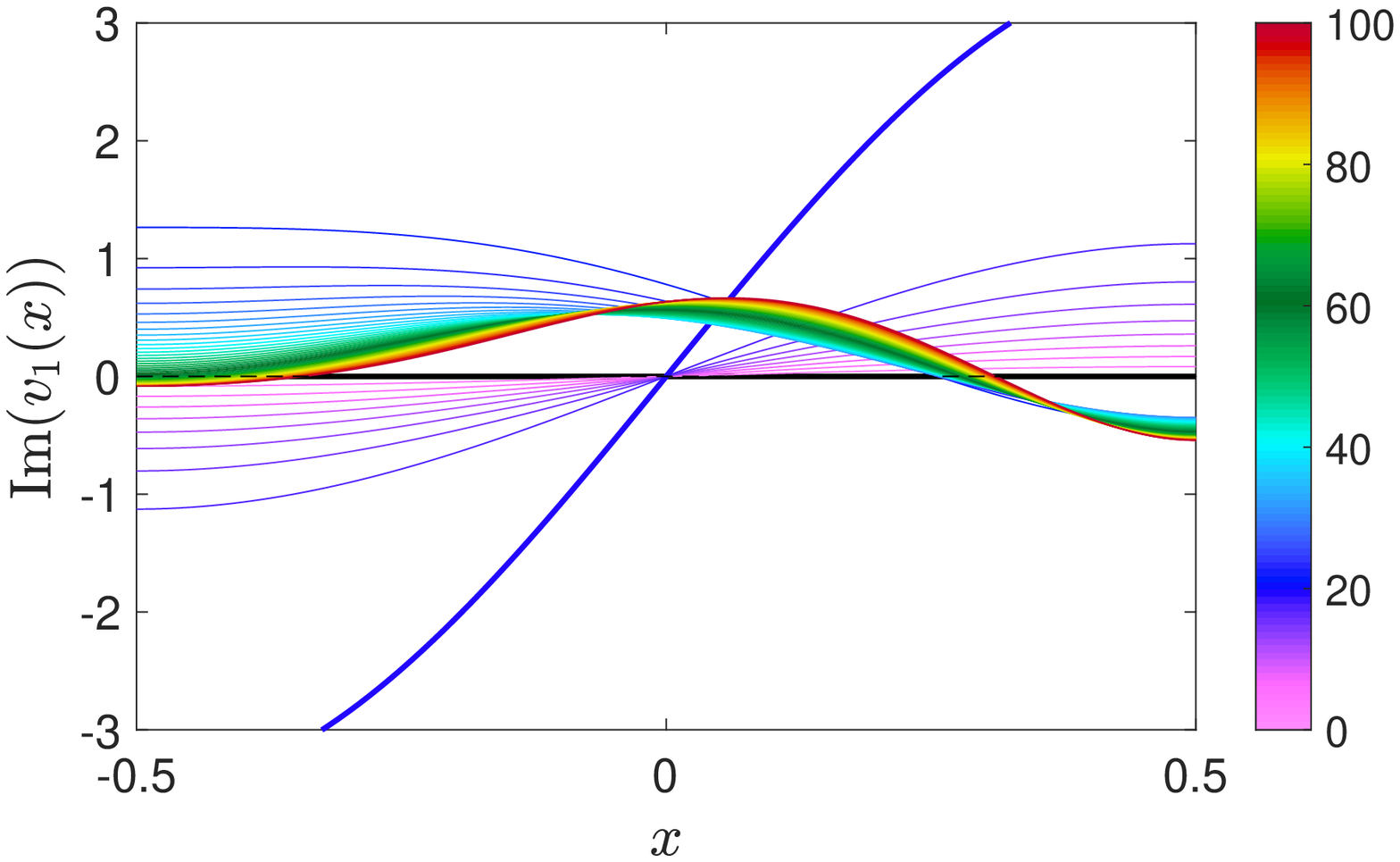} % {1d_v1_Im.eps}
\includegraphics[width=0.49\linewidth]{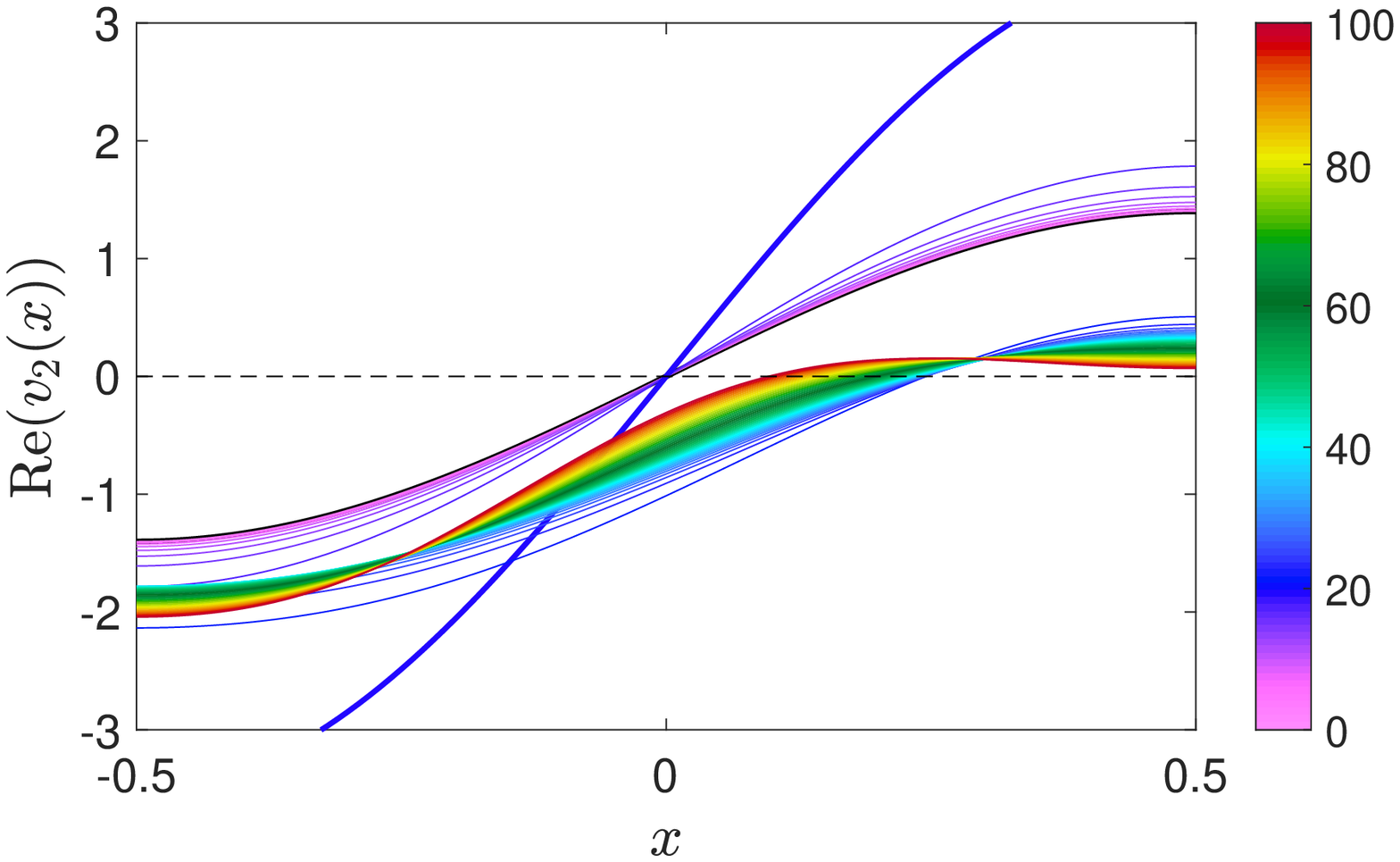} % {1d_v2_Re.eps}
\includegraphics[width=0.49\linewidth]{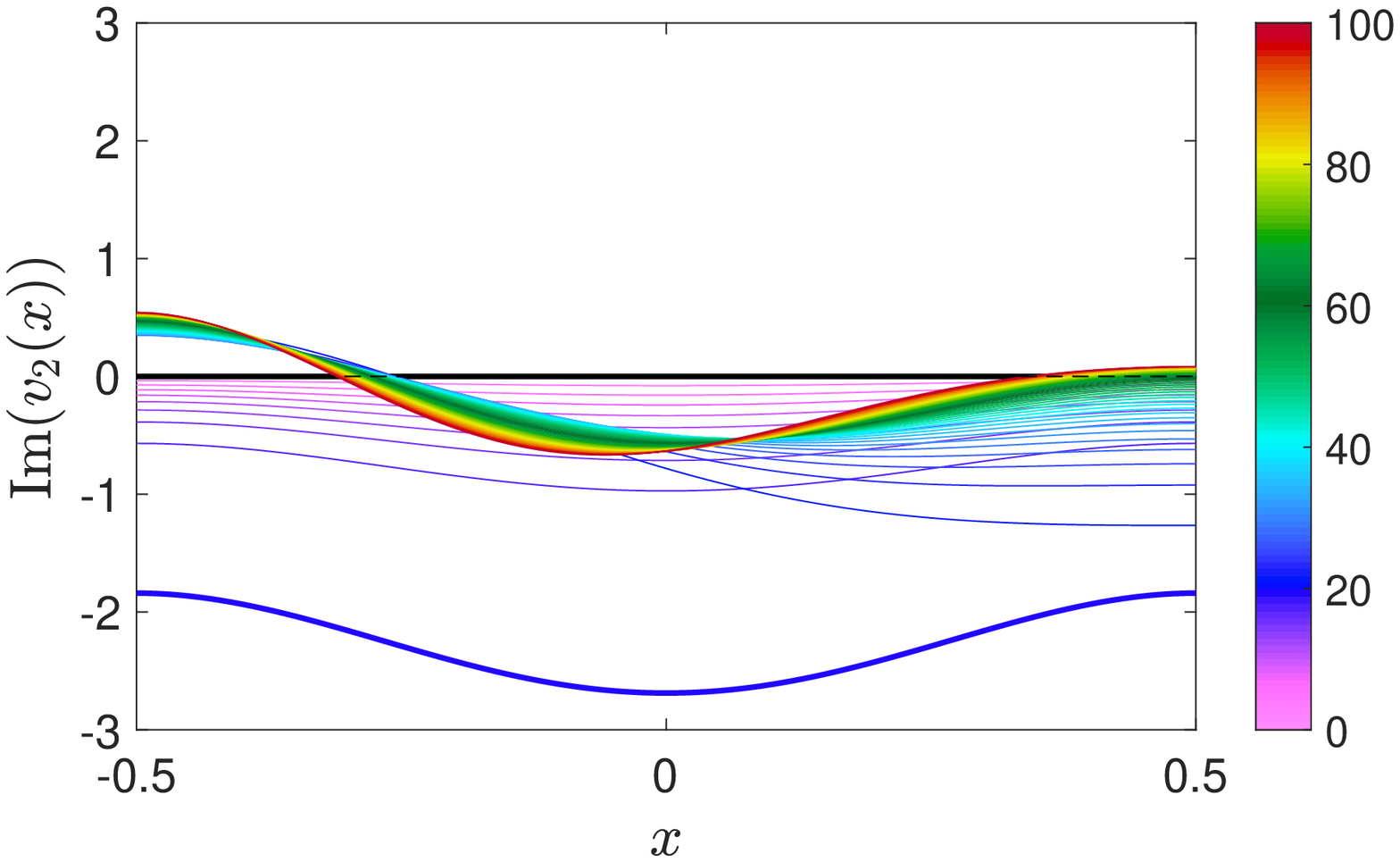} % {1d_v2_Im.eps}
\caption{
The spectrum of the Bloch-Torrey operator on the interval $(-1/2,1/2)$
with Neumann boundary conditions.  (Top) The real (left) and imaginary
(right) part of the first two eigenvalues $\lambda_1(g)$ and
$\lambda_2(g)$ as functions of $g$.  A transition from of two real
eigenvalues to a complex conjugate pair occurs at the branch point
$g_c \approx 18.06$.  (Middle, Bottom) The real (left) and imaginary
(right) part of the first eigenmode $v_1(x)$ (middle) and of the
second eigenmode $v_2(x)$ (bottom).  $51$ plotted curves correspond to
$51$ equally spaced values of $g$ from $0$ to $100$ (shown by
colorbar).  Black line depicts the Laplacian eigenmodes ($g = 0$),
whereas thick blue line shows the eigenmodes at $g = 18$ (near the
branch point $g_c \approx 18.06$). }
\label{fig:1D}
% [lam,v1,v2] = A_Moutal_article_1d_eigen2(lam,v1,v2);
\end{figure}

Figure \ref{fig:1D} illustrates their discovery and the above
discussion for the unit interval $(-1/2,1/2)$.  On the top, the real
and imaginary parts of the first two eigenvalues $\lambda_1(g)$ and
$\lambda_2(g)$ are shown as functions of $g$ (this figure can be
considered as a zoom of the rescaled figure 3 from
\cite{Stoller1991a}).  At $g = 0$, one retrieves the eigenvalues of
the second derivative operator $\B(0) = -d^2/dx^2$: $\lambda_n(0) =
\pi^2 (n-1)^2$, with $n = 1,2,3,\ldots$.  A transition from two
real eigenvalues to a complex conjugate pair occurs at the branch
point $g_c \approx 18.1$, in perfect agreement with the theoretical
prediction by Stoller {\it et al.} (see also
Sec. \ref{sec:discussion}).  This qualitative change of the
eigenvalues is also reflected in the behavior of the associated
eigenmodes $v_1(x)$ and $v_2(x)$, whose real and imaginary parts are
shown on the middle and bottom panels of Fig.  \ref{fig:1D}.  Here we
plot $51$ eigenmodes for $51$ equally spaced values of $g$ ranging
from $0$ to $100$.  At $g = 0$, one retrieves the eigenmodes of the
second derivative with Neumann boundary conditions at $x = \pm 1/2$,
namely $v_1(x) = 1$ and $v_2(x) = \sqrt{2}\sin(\pi x)$.  As $g$
increases up to $g_c$, the shape of these eigenmodes continuously
changes but they remain to be qualitatively similar and
``delocalized''.  In particular, the real parts of $v_1(x)$ and
$v_2(x)$ are respectively symmetric and antisymmetric with respect to
the middle point of the interval.
However, at the branch point, there is a drastic change in the
behavior of these eigenmodes.  As two eigenmodes coalesce into a
single one, their normalization factors $(v_1|v_1)^2$ and
$(v_2|v_2)^2$ vanish (see below), implying a considerable increase of
the amplitudes of $v_1$ and $v_2$ for $g$ near $g_0$ (see thick blue
curves).  For $g > g_c$, the parity symmetry of these eigenmodes
implies $\mathrm{Re}(v_2(x)) = \mathrm{Re} (v_1(1-x))$ and
$\mathrm{Im}(v_2(x)) = - \mathrm{Im} (v_1(1-x))$.  As $g$ increases,
the eigenmode $v_1(x)$ progressively shifts to the right endpoint $x =
1$, near which it is getting more and more localized (i.e., its
amplitude on the left half of the interval is reduced).  Likewise, the
eigenmode $v_2(x)$ is getting localized near the left endpoint $x =
0$.  Such an abrupt change in the shape and symmetries of the
eigenmodes at the branch point is a remarkable feature of
non-Hermitian operators.

The above argument also implies that if the branch point $g_0$ with
the smallest amplitude $|g_0|$ lies on the real axis, all eigenvalues
$\lambda_n(g)$ of the Bloch-Torrey operator are necessarily real for
any $0 < g < |g_0|$.  This is an example of a non-Hermitian operator
with a real spectrum.  Actually, there are many families of such
operators
\cite{Bender1998a,Cannata1998a,Delabaere1998a,Delabaere1998b,Bender1999a,Fernandez1999a,Mezincescu2000a,Delabaere2000a}
(see also a review \cite{Bender2007a}).

\subsection{Arbitrary bounded domains}

For a parity-symmetric domain, the spectrum exhibits branch points at
particular values of the gradient $g \in \R$.  In contrast, for a
general domain without parity symmetry, all eigenvalues are generally
complex for non-zero $g$ and there is generally no branch point on the
real line.  However, this is true only for real values of $g$.  By
allowing $g$ to take complex values, one may recover branch points for
general bounded domains.  In other words, an asymmetry of the domain
typically ``shifts'' branch points away from the real axis.

\begin{figure}[tp]
\centering
\includegraphics[width=0.99\linewidth]{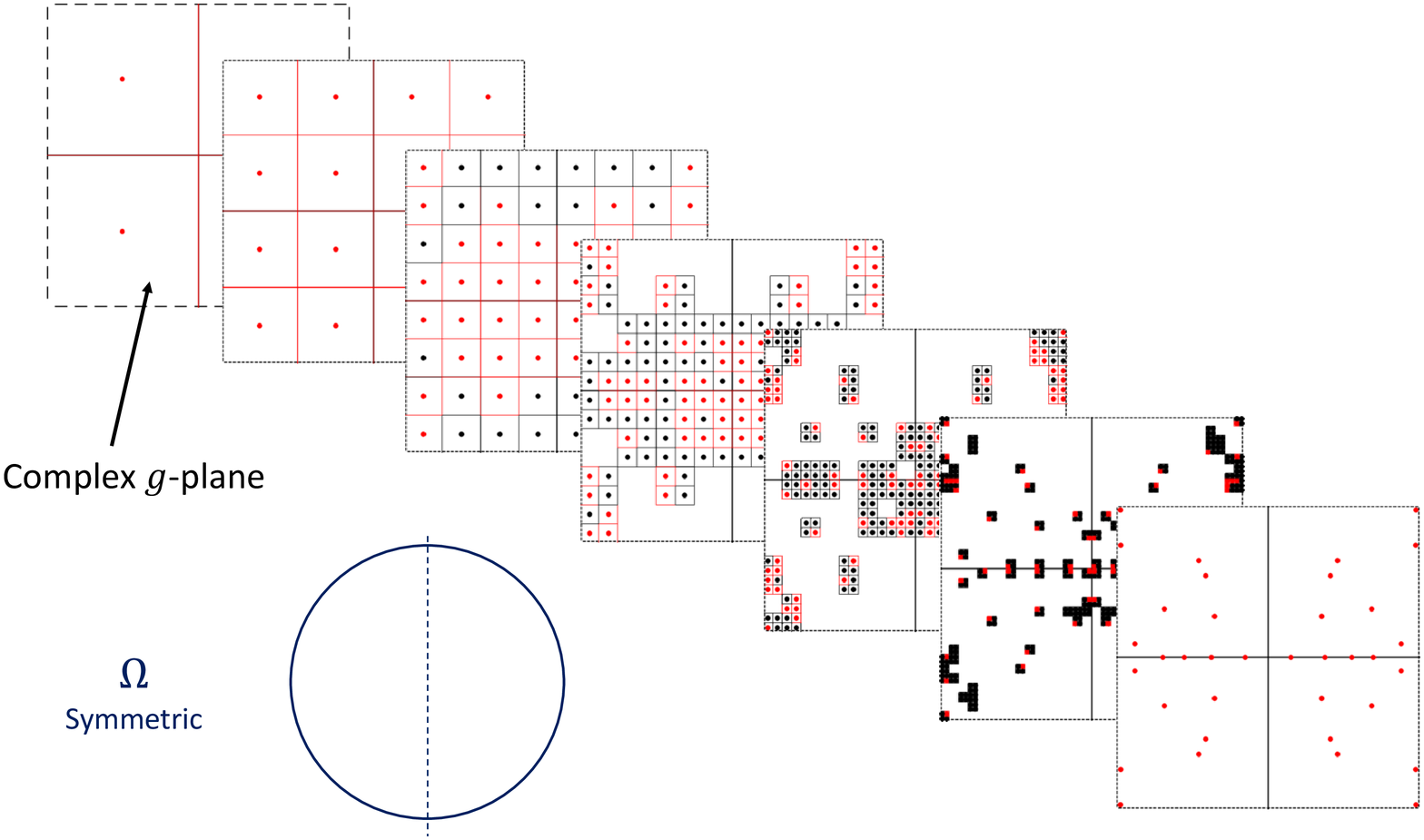} % bifurcation_complexe_disk_new_dg.pdf}
\noindent\rule[5pt]{\linewidth}{0.4pt}
\includegraphics[width=0.99\linewidth]{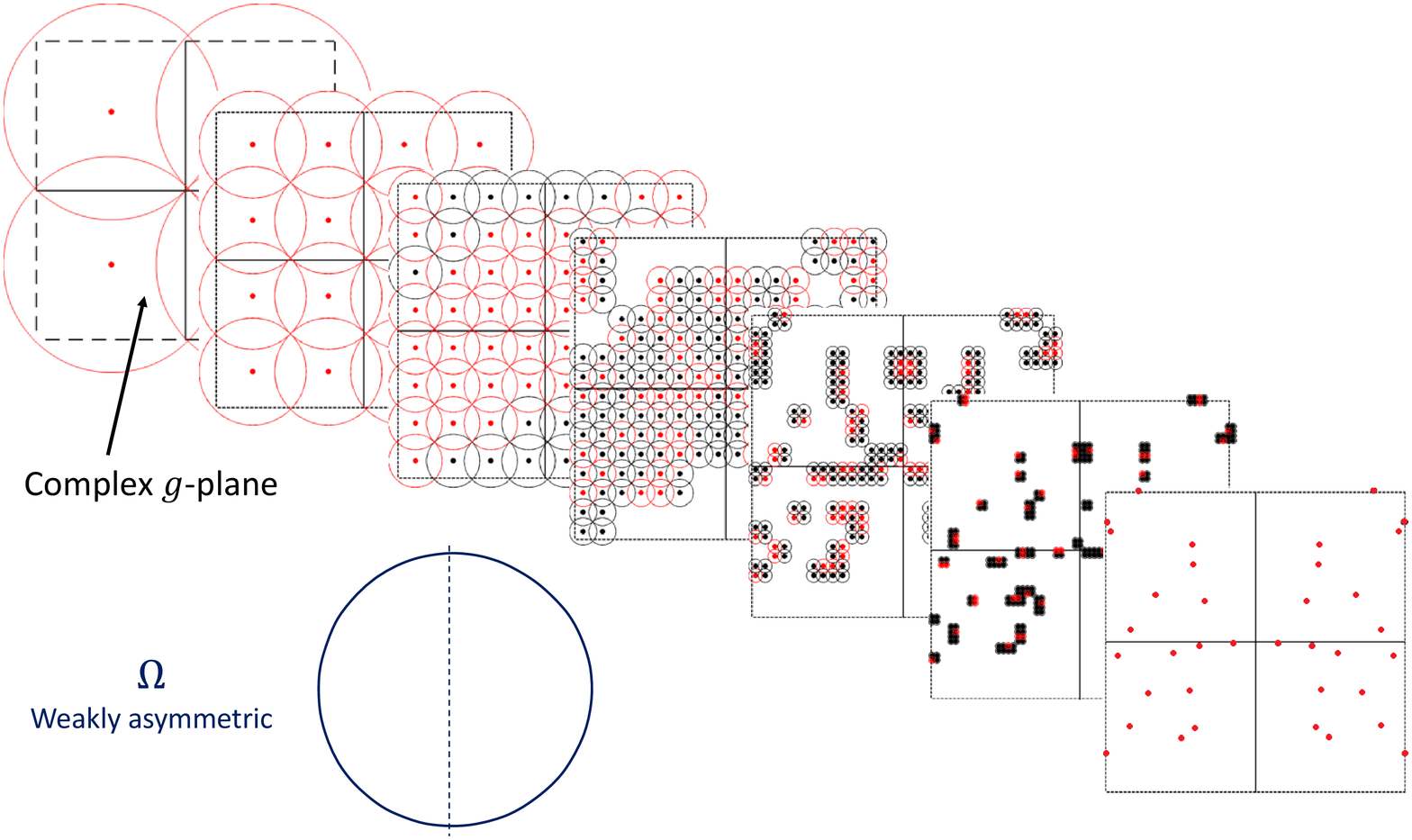} % bifurcation_complexe_asym_dg.pdf}
\caption{
Several iterations of the algorithm that finds spectral branch points
of the Bloch-Torrey operator in the complex $g$-plane.  The red dots
indicate contours that yield a non-zero contour integral.  The initial
range of $g$ is a square in the complex plane, from $-200(1+i)$ to
$200 (1+i)$.  We emphasize that there are infinitely many branch
points in the complex plane but only a finite number appears because
of the finite range of $g$.  (Top) The domain $\Omega$ is a disk and
we apply the algorithm with square integration contours.  (Bottom) The
domain $\Omega$ is slightly asymmetric (a thin dashed line helps to
visualize the asymmetry) and we apply the algorithm with circular
integration contours.  Compared to the pattern of branch points for a
disk, the top-bottom symmetry is lost but branch points still exist.
Note that the left-right symmetry is preserved in both plots, as
discussed in the text.}
\label{fig:branchement_complexe_disk}
\end{figure}

The point of view onto spectral branch points as complex branch points
of a multi-valued function reveals a way to find them in the complex
plane.  Let us consider a closed contour $\mathcal{C}_g$ in the
complex $g$-plane and compute the contour integral
\begin{equation}
I_n(\mathcal C_g) = \oint_{\mathcal C_g}\lambda_n(g) \,\mathrm{d}g\;.
\end{equation}
If $\mathcal C_g$ does not enclose any branch point, then
$\lambda_n(g)$ is analytical inside the contour for any sheet
$n=1,2,\ldots$, and one has $I_n(\mathcal C_g) = 0$.  In contrast, if
the contour $\mathcal{C}_g$ encloses branch points, the contour
integral along $\mathcal{C}_g$ is generally non zero anymore.
Therefore, one can find branch points by the following algorithm:
\begin{enumerate}
\item 
choose an initial closed contour $\mathcal{C}_g$ and compute
$I_n(\mathcal C_g)$ with $n=1,2,\ldots,N$ for a large enough number
$N$ of sheets by following continuously the contour $\mathcal C_g$;

\item 
if the obtained value is not zero, split the contour in smaller closed
contours and perform the integral over each smaller contour;

\item 
identify contours with non-zero integrals and repeat the previous step
for each of them.
\end{enumerate}
In this way, one can determine a finite number of branch points inside
the initial contour.  However, additional restrictions on the spectral
properties are needed to ensure that {\it all} branch points inside
the contour are identified.  For instance, branch points should not
accumulate near a point inside the contour.
Figure \ref{fig:branchement_complexe_disk} illustrates an application
of this algorithm for the Bloch-Torrey operator on a symmetric domain
(a disk) and an asymmetric domain (an oval).

In practice, each integral is computed numerically by discretizing its
contour $\mathcal{C}_g$, i.e., $I_n(\mathcal C_g) \approx I_n^{\rm
num}(\mathcal C_g)$.  As a consequence, the numerically computed
integral is never equal to $0$, even if there is no branch point
inside the contour.  One needs therefore to choose a numerical
threshold $\chi$ to distinguish between ``zero'' and ``non-zero''
integrals (indicated in Fig. \ref{fig:branchement_complexe_disk} by
black and red dots, respectively).  A compromise should be found
between reliability and speed of the algorithm.  In fact, a too high
threshold may result in missed branch points: if $0< |I_n(\mathcal
C_g)| < \chi$, such branch point(s) inside the contour $\mathcal C_g$
are not detected.  In turn, a too small threshold generally leads to a
large number of evaluations of contour integrals: even if there is no
branch point inside $\mathcal C_g$ (and thus $I_n(\mathcal C_g) = 0$),
numerical errors may result in $|I_n^{\rm num}(\mathcal C_g)| > \chi$,
so that the algorithm continues subdivisions into smaller contours
until all the contour integrals become less than $\chi$.  Since each
integral requires the computation of the eigenvalues $\lambda_n(g)$
along the contour $\mathcal{C}_g$, a bad choice of the threshold may
be very time-consuming (see \ref{sec:numerics} for the spectrum
computation in the case of the Bloch-Torrey operator).  For the
particular example shown on Fig. \ref{fig:branchement_complexe_disk},
one can see that the threshold was chosen somewhat too low because
some red squares at the initial steps eventually disappeared after a
large number of iterations.  In other words, a suspicion of missed
branch points was dismissed but it required an excessive number of
computations.

The choice of the contour is also a matter of compromise.  On one
hand, one can choose a contour shape that tiles the plane, like a
square.  This choice allows one to have non-overlapping integration
contours to avoid counting one branch point twice.  On the other hand,
one can choose a smooth contour, like a circle.  This results in a
higher numerical accuracy for the integral computation and thus allows
one to lower the threshold.  However, this requires an additional
criterion to discard ``double'' branch points that may result from
overlapping contours.  Both options are illustrated on
Fig. \ref{fig:branchement_complexe_disk}.
 
\begin{figure}[tp]
\centering
\includegraphics[width=0.99\linewidth]{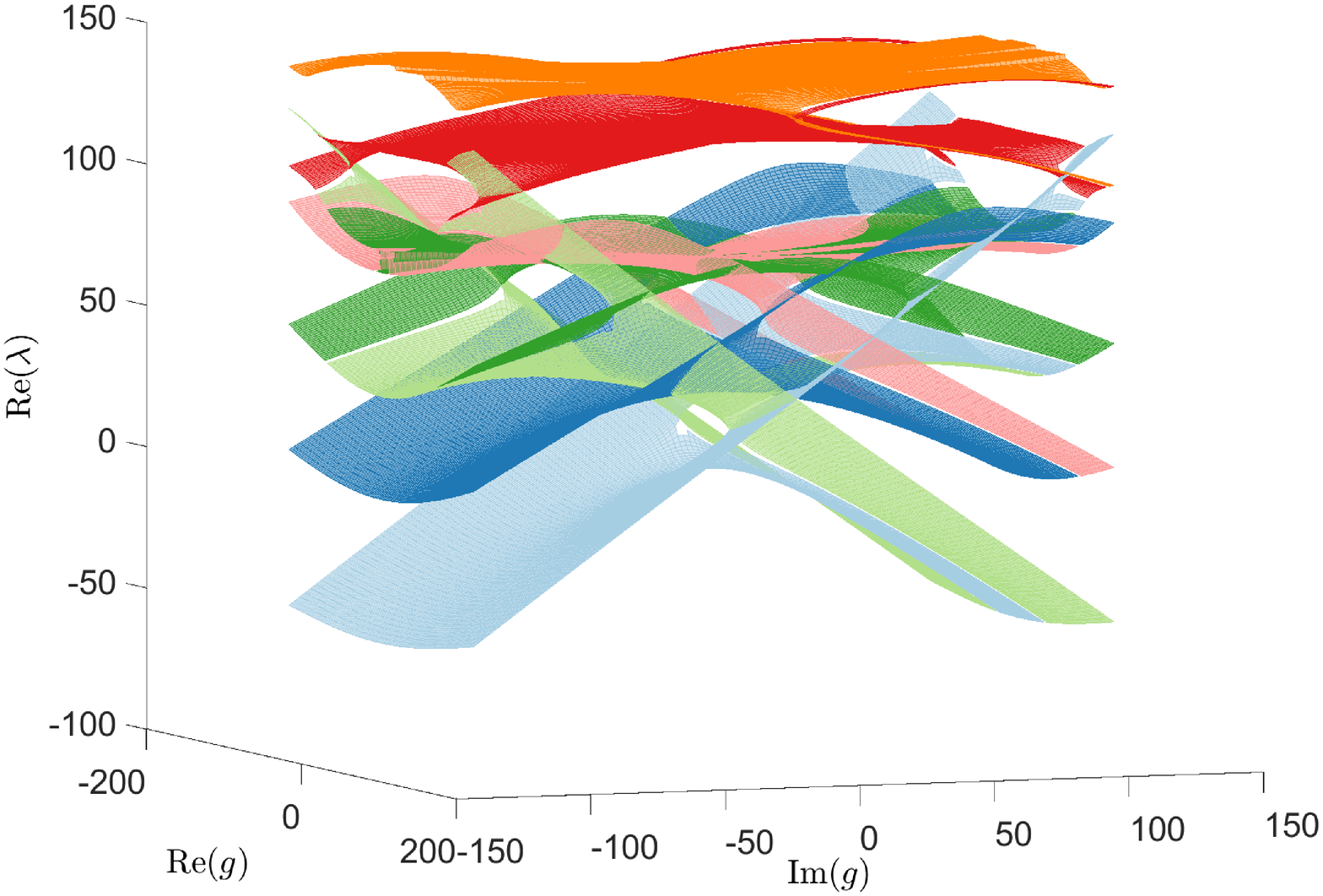} % feuilles_spectre_real_dg.eps}
\includegraphics[width=0.99\linewidth]{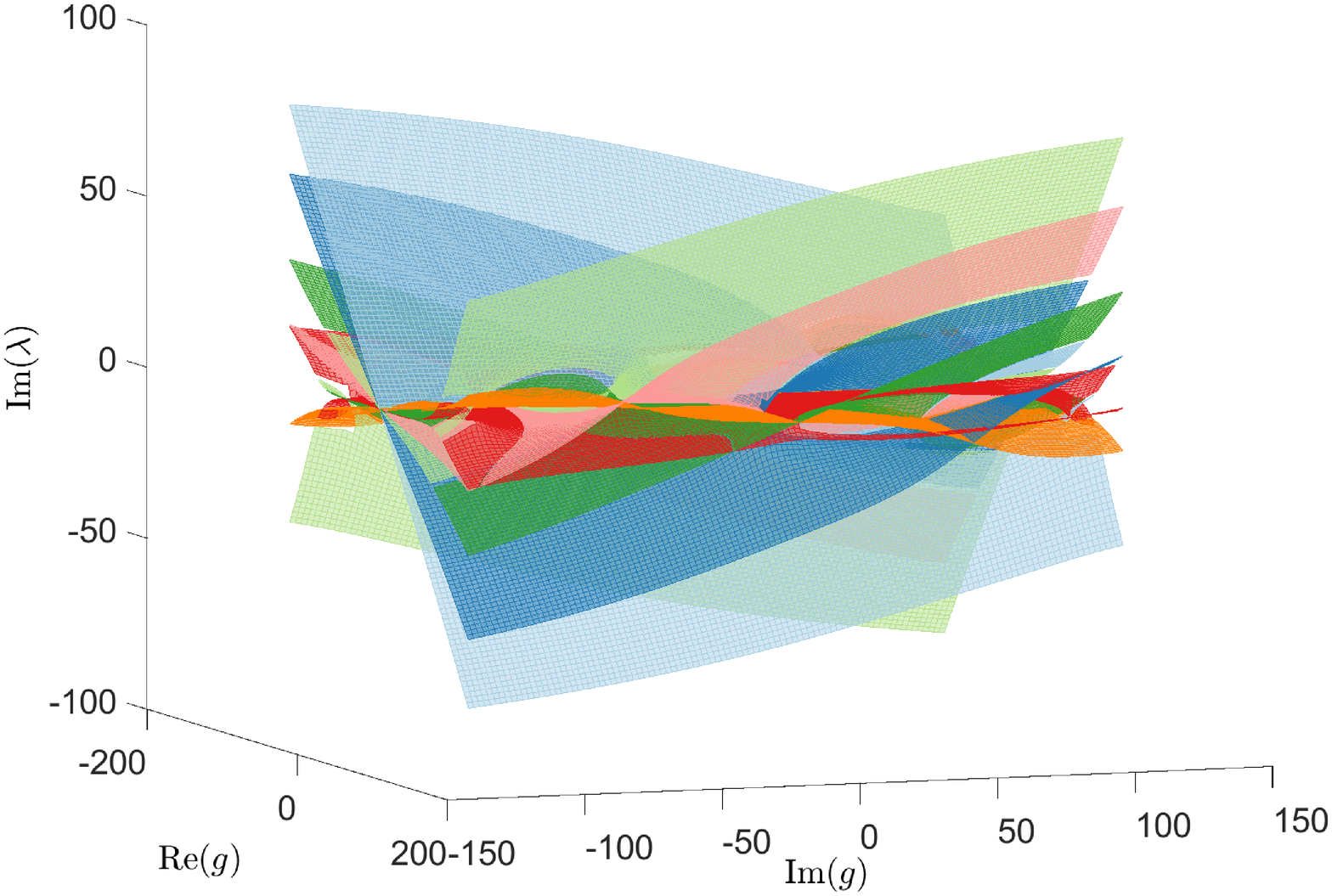} % feuilles_spectre_imag_dg.eps}
\caption{
Real (top) and imaginary (bottom) part of the spectrum of the
Bloch-Torrey operator in a disk (see also top panel of
Fig. \ref{fig:branchement_complexe_disk}).  Several sheets are shown
by different colors.  The figure reveals a rich pattern of branch
points, with a similar structure as in
Figs. \ref{fig:branchement_racine} and
\ref{fig:branchement_complexe}.}
\label{fig:feuilles_spectre}
\end{figure}

Let us inspect the pattern of branch points shown on
Fig. \ref{fig:branchement_complexe_disk}.  Performing a complex
conjugation of the Bloch-Torrey operator,
\begin{equation}
\mathcal{B}(g)^* = -\nabla^2 + i g^* x = -\nabla^2 - i (-g^*) x\;,
\end{equation}
we see that the branch points are always symmetric under the
transformation $g \to -g^*$, that explains the left-right symmetry of
their pattern.  Furthermore, the pattern for $x$-parity-symmetric
domains (like a disk) exhibits a top-bottom symmetry, according to
\begin{equation}
\mathcal{P}_x \mathcal{B}(g)^* = -\nabla^2 - i g^* x\;,
\end{equation}
where we used the parity symmetry to write $\mathcal{P}_x \nabla^2 =
\nabla^2$.  The above equation shows that for an $x$-parity symmetric
domain, the pattern of branch points is symmetric under the
transformation $g \to g^*$, i.e. the top-bottom symmetry.  Note that
the existence of branch points on the real axis is consistent with the
top-bottom symmetry of their pattern.  

To get a complementary insight, Fig. \ref{fig:feuilles_spectre} shows
several sheets of the multi-valued function $\lambda(g)$ for the
Bloch-Torrey operator in a disk.  Although the figure is visually
complicated due to the superposition of numerous sheets, one can
recognize the basic square-root structure of branch points illustrated
on Fig. \ref{fig:branchement_racine}.

\subsection{Spectral decomposition at a branch point}
\label{section:app_bifurcation_spectral_decomposition}

Let us ``translate'' the general statements of
Sec. \ref{section:bifurcation_matrix} about the second-order branch
points analyzed with a matrix model, into the language of eigenmodes
of the Bloch-Torrey operator.  In particular, we investigate the
validity of the spectral decomposition
\begin{equation}
f(\mathbf r) \overset{?}{=} \sum_n {(f|v_n)} v_n(\mathbf r)\;,
\label{eq:app_BT_projection}
\end{equation}
where the bilinear form $(\cdot|\cdot)$ is defined in
Eq. \eqref{eq:scalar}.  As the eigenbasis formed by $v_n$ is not
necessarily complete, the validity of
Eq. (\ref{eq:app_BT_projection}), which was often used in physics
literature, is not granted.  As the arguments and techniques
employed below are rather standard, we focus on qualitative
explanations while mathematical details can be found, e.g., in
\cite{Seyranian,Seyranian05,Kirillov05,Gunther06,Gunther2007a}.

\subsubsection{Behavior of the eigenmodes at a branch point.}

Let us consider two eigenpairs $(v_1,\lambda_1)$ and $(v_2,\lambda_2)$
that undergo branching at $g = g_0$.  The matrix model of
Sec. \ref{section:bifurcation_matrix} shows that $v_1$ and $v_2$
collapse onto a single eigenmode $v_0$ at the branch point.  Moreover,
since $v_1$ and $v_2$ are ``orthogonal'' with respect to the bilinear
form $(\cdot | \cdot)$ if $g \neq g_0$, we conclude by continuity that
$v_0$ is self-orthogonal, i.e. $(v_0 | v_0)=0$.  We outline that the
condition $(v_0 | v_0)=0$ may be achieved for a nonzero complex-valued
function $v_0$ because $(\cdot | \cdot)$ does not contain complex
conjugate, see Eq. (\ref{eq:scalar}).

The computations in Sec. \ref{section:bifurcation_matrix} imply the
following behavior in a vicinity of the branch point $g_0$:
\begin{subequations}  \label{eq:v1v2}
\begin{eqnarray}
&v_1(\mathbf r) \approx \beta_1(g) \left[v_0(\mathbf r) + (g - g_0)^{1/2} y_0(\mathbf r)\right] \;, \\
& v_2(\mathbf r) \approx \beta_2(g) \left[v_0(\mathbf r) - (g - g_0)^{1/2} y_0(\mathbf r)\right]\;,
\end{eqnarray}
\end{subequations}
where $\beta_1(g)$ and $\beta_2(g)$ are normalization constants, and
the function $y_0(\mathbf r)$ is \textit{a priori} unknown and depends
on the details of the branch point under study.  The form
(\ref{eq:v1v2}) ensures that $v_1$ and $v_2$ are orthogonal to each
other to the first order in $(g - g_0)^{1/2}$, i.e., $(v_1|v_2) =
O(g-g_0)$.  Moreover, the normalization conditions imply
\begin{subequations}
\begin{eqnarray}
1 & = (v_1|v_1) \approx 2\beta_1^2(g) (g - g_0)^{1/2} (v_0 | y_0) , \\
1 & = (v_2|v_2) \approx -2\beta_2^2(g) (g - g_0)^{1/2} (v_0 | y_0) ,
\end{eqnarray}
\end{subequations}
therefore
\begin{subequations}  \label{eq:v1v2_bis}
\begin{eqnarray}
&v_1(\mathbf r) \approx C (g - g_0)^{-1/4} v_0(\mathbf r) + C (g - g_0)^{1/4} y_0(\mathbf r) \;, \label{eq:app_BT_mode_bifurc1}\\
& v_2(\mathbf r) \approx iC (g - g_0)^{-1/4} v_0(\mathbf r) - iC (g - g_0)^{1/4} y_0(\mathbf r)\;, \label{eq:app_BT_mode_bifurc2}
\end{eqnarray}
\end{subequations}
with the constant $C=(2(v_0 | y_0))^{-1/2}$.  We recall that the
eigenvalues $\lambda_1$ and $\lambda_2$ behave near $g_0$ as
\begin{equation}
\lambda_1 \approx \lambda_0(g_0) + (g - g_0)^{1/2} \eta_0 \;, \qquad \lambda_1 \approx \lambda_0(g_0) - (g - g_0)^{1/2} \eta_0 \;,
\end{equation}
with an unknown coefficient $\eta_0$. By writing the Bloch-Torrey
operator as
\begin{equation}
\mathcal{B} = \mathcal{B}_0 - i(g - g_0)x\;,  \qquad \mathrm{with} \quad \mathcal{B}_0 = -\nabla^2 - i g_0 x,
\end{equation}
one can expand the eigenmode equation $\mathcal{B} v_1 = \lambda_1
v_1$ in powers of $(g - g_0)$ and get in the lowest order
\begin{equation}
\mathcal{B}_0 y_0 = \lambda_0 y_0 + \eta_0 v_0\;.
\label{eq:app_Jordan_block}
\end{equation}
One recognizes in this equation the typical Jordan block associated to
a branch point (see Sec. \ref{section:bifurcation_matrix}).

\subsubsection{Regularity of the spectral decomposition at a branch point.}

The above equations (\ref{eq:v1v2_bis}) reveal that the eigenmodes
$v_1$ and $v_2$ diverge as $(g - g_0)^{-1/4}$ at the branch point.
This behavior is intuitively expected because $v_1$ and $v_2$ tend to
the self-orthogonal eigenmode $v_0$, therefore the normalization
constants $\beta_1$ and $\beta_2$ should diverge as $g \to g_0$. One
may wonder whether this divergence produces specific effects in the
spectral decomposition \eqref{eq:app_BT_projection} such as a
resonance effect when two eigenmodes near a branch point could
dominate the series.  We show here that this is not the case because
two infinitely large contributions in Eq. \eqref{eq:v1v2_bis} cancel
each other, yielding a continuous behavior in the limit $g \to g_0$.
Note that this regularization follows from the general argument that
the projector $\Pi(g)$ over the space spanned by $v_1$ and $v_2$ is an
analytical function of $g$ at the branch point (see
Sec. \ref{section:bifurcation_matrix}).

Let us isolate the terms with $v_1$ and $v_2$ in the sum
\eqref{eq:app_BT_projection} and define
\begin{equation}
f_{1,2}(\mathbf r)= (f|v_1) v_1(\mathbf r) + (f|v_2) v_2(\mathbf r)\;.
\end{equation}
Now we use the expansions \eqref{eq:v1v2_bis} to obtain, in a vicinity
of the branch point:
\begin{eqnarray}
f_{1,2}(\mathbf{r}) &\approx C^2 \biggl((g - g_0)^{-1/4} (f| v_0) + (g- g_0)^{1/4}  (f|y_0)\biggr) \nonumber\\
&\times \biggl((g - g_0)^{-1/4} v_0(\mathbf r) + (g - g_0)^{1/4} y_0(\mathbf r)  \biggr)\nonumber\\
&-C^2 \biggl((g - g_0)^{-1/4} (f| v_0) - (g - g_0)^{1/4}  (f|y_0)\biggr)\nonumber \\
&\times\biggl((g - g_0)^{-1/4} v_0(\mathbf r) - (g - g_0)^{1/4} y_0(\mathbf r)  \biggr),
\end{eqnarray}
which simplifies to
\begin{equation}
f_{1,2}(\mathbf r) \approx \frac{(f|y_0)}{(v_0|y_0)} v_0(\mathbf r) + \frac{(f|v_0)}{(v_0|y_0)} y_0(\mathbf r)
+ O((g - g_0)^{1/2})\;.
\end{equation}

Two important observations can be made: (i) the diverging terms $(g -
g_0)^{-1/4}$ have canceled each other so that $f_{1,2}(\mathbf r)$ has
a finite value in the limit $g \to g_0$; (ii) at the branch point,
$f_{1,2}(\mathbf r)$ is expressed as a linear combination of the
eigenmode $v_0(\mathbf r)$ and the additional function $y_0(\mathbf
r)$. This shows that the spectral decomposition is still valid if the
eigenmode family is supplemented with a ``generalized eigenmode''
$y_0(\mathbf r)$. Note that the function $y_0(\mathbf r)$ is the
analogous of the vector $Y_0$ for the matrix model considered in
Sec. \ref{section:bifurcation_matrix}.

If the function $f$ represents the magnetization, then one can compute
its time evolution by exponentiating the Bloch-Torrey operator
$\B(g_0)$ over the basis $(v_0,y_0,v_3,v_4,\ldots)$.  The only
difference with the general case $g\ne g_0$ lies in the $2\times 2$
Jordan block associated to the pair $v_0,y_0$ (see
Eq. \eqref{eq:app_Jordan_block}) that yields
\begin{equation}
\exp\left(-t \left[\begin{array}{cc} \lambda_0 & \eta_0 \\ 0 & \lambda_0\end{array}\right]\right) 
= \exp(-\lambda_0 t) \left[\begin{array}{cc} 1 & - \eta_0 t \\ 0 & 1 \end{array} \right]\;.
\end{equation}
One may recognize the typical $te^{-t}$ evolution of a critically
damped harmonic oscillator, which also originates from the exponential
of a Jordan block.  Therefore, the evolution of the magnetization
during an extended gradient pulse at $g_0$ is given by
\begin{equation}
\fl
m(t,\mathbf r) = \frac{(1|y_0) - \eta_0 t (1|v_0)}{(v_0|y_0)} v_0(\mathbf r) e^{-\lambda_0 t} 
+ \frac{(1|v_0)}{(v_0|y_0)} y_0(\mathbf r) e^{-\lambda_0 t} 
+\sum_{n\geq 3} (1|v_n) v_n(\mathbf r) e^{-\lambda_n t}\;.
\end{equation}
 
In summary, the spectral expansion \eqref{eq:app_BT_projection} is
valid for any $g$ except for a set $\mathcal E$ of branch points.  In
turn, when $g = g_0 \in \mathcal E$, the two eigenmodes $v_n$ and
$v_{n'}$ that correspond to equal eigenvalues $\lambda_n(g_0) =
\lambda_{n'}(g_0)$, merge into a single eigenmode $v_0$ that should thus
be supplemented by a generalized eigenmode $y_0$.  A rigorous
demonstration of this conclusion and its further developments present
an interesting perspective.

\subsection{Summary}

\begin{figure}[tp]
\centering
\includegraphics[width=0.9\linewidth]{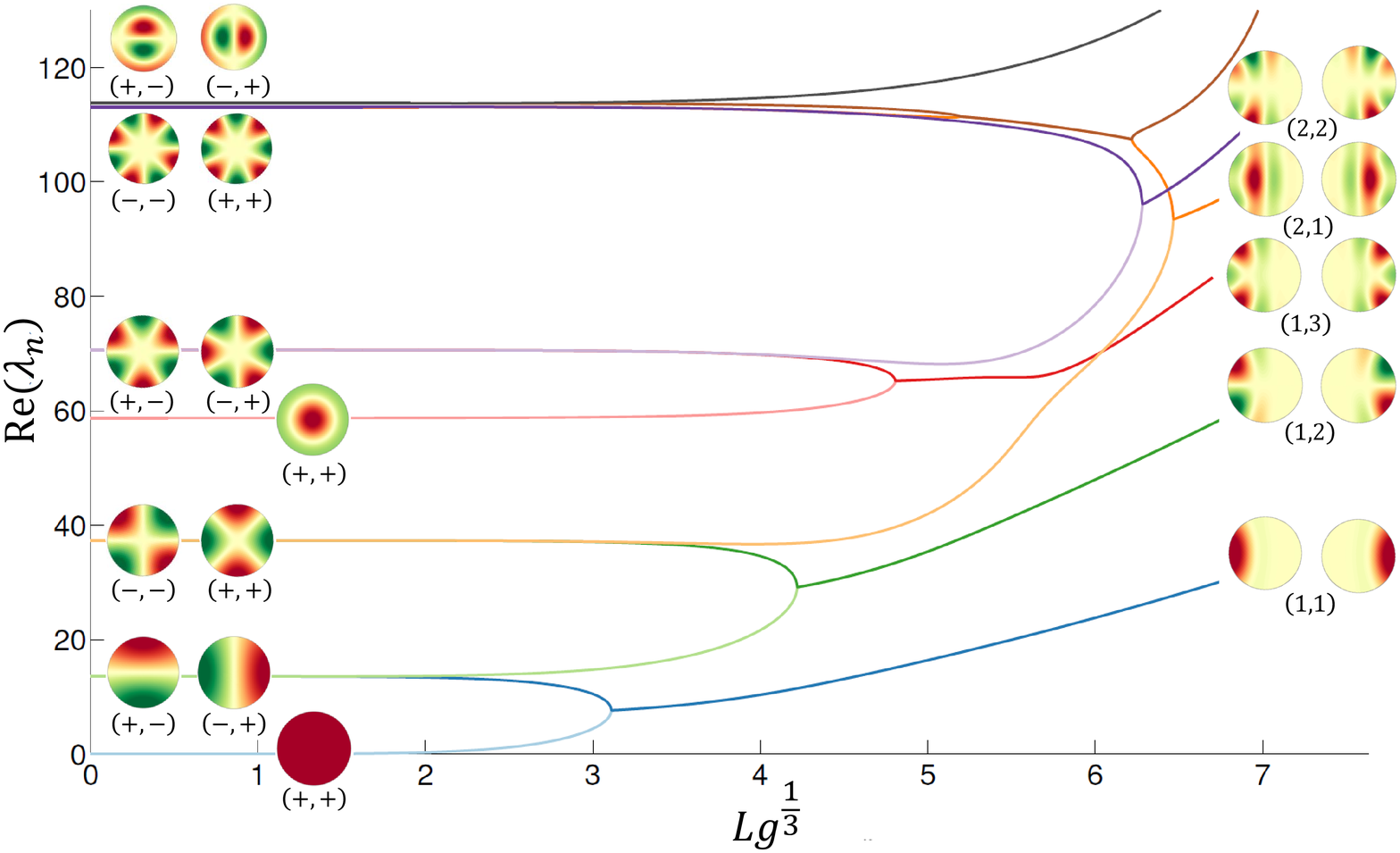} % bifurcation_disk_real4_dg.pdf}
\includegraphics[width=0.9\linewidth]{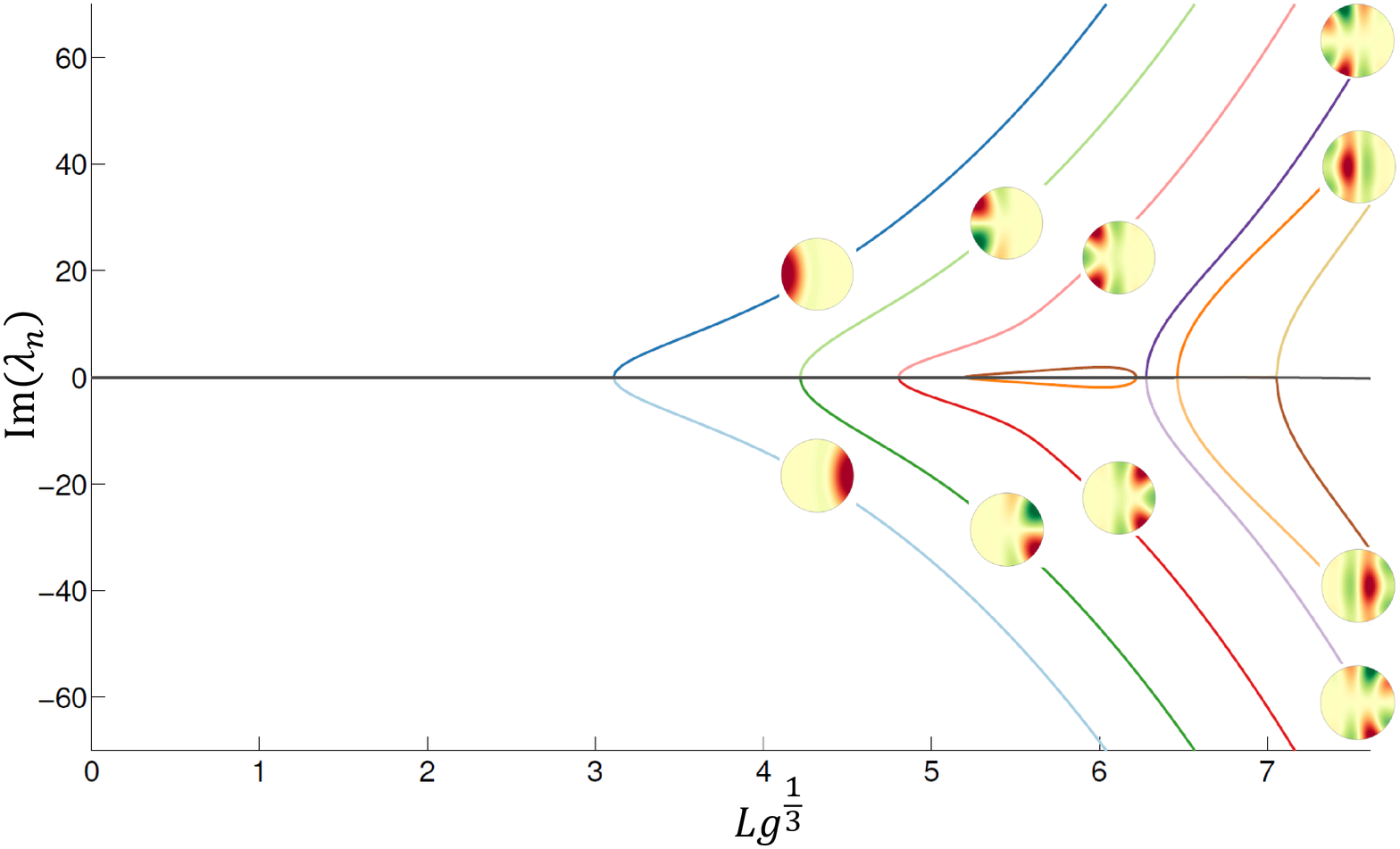} % bifurcation_disk_imag4_dg.pdf}
\caption{
Spectral properties of the Bloch-Torrey operator in a disk of diameter
$L = 1$.  Real (top) and imaginary (bottom) parts of several
eigenvalues $\lambda_n(g)$ as functions of $L g^{1/3}$.  At $g = 0$,
one retrieves the eigenvalues of the Laplace operator with Neumann
boundary condition: $\lambda_1(0) = 0$, $\lambda_2(0) =
\lambda_3(0) = 4(j'_{1,1})^2 \approx 13.56$, $\lambda_4(0) =
\lambda_5(0) = 4(j'_{2,1})^2 \approx 37.31$, $\lambda_6(0) =
4(j'_{0,2})^2 \approx 58.68$, etc., where $j'_{n,k}$ denotes the
$k$-th zero of $J'_n(z)$ \cite{Grebenkov2013b}.  The symmetries of the
corresponding eigenmodes are distinguished by two signs $(\pm,\pm)$.
The spectrum exhibits branch points for some values of $g$, above
which the corresponding eigenmodes drastically change their shapes.
For $g > g_c$, the profile of the corresponding eigenmodes is
characterized by two indices $(n,l)$ that represent loosely the number
of oscillations perpendicular and parallel to the boundary,
respectively. }
\label{fig:pure_beauty}
\end{figure}

Figure \ref{fig:pure_beauty} summarizes our findings by showing
several eigenvalues and the corresponding eigenmodes of the
Bloch-Torrey operator in a disk of diameter $L$, as a function of the
dimensionless quantity $L g^{1/3}$.  The power $1/3$ has no particular
significance but was chosen to improve the clarity of the figure.  At
$g = 0$, the Bloch-Torrey operator is reduced to the Laplace operator
with the well-known eigenmodes \cite{Grebenkov2013b}.  The rotational
invariance of the disk implies that some eigenvalues are twice
degenerate, in which case one eigenmode can be transformed to its
``twin'' by an appropriate rotation.
When $g > 0$ is small enough, the symmetries of the eigenmodes are
preserved and can be distinguished by signs $(\pm,\pm)$, where the
first sign refers to the symmetry along the $x$-axis ($+$ for
symmetric and $-$ for antisymmetric), and the second sign refers to
the symmetry along the $y$-axis.  The symmetry of the Laplacian
eigenmodes is of considerable importance because the gradient term $ig
x$ couples only the modes with the same symmetry along $y$ and with
the opposite symmetries along $x$.  In other words, the gradient
couples $(+,+)$ to $(-,+)$ and $(-,-)$ to $(+,-)$.  For example, the
first branch point (blue curves) involves the constant eigenmode with
symmetry $(+,+)$, and the eigenmode with symmetry $(-,+)$ immediately
above.  A more complicated branching pattern can be observed with the
light orange curve, which has a $(+,+)$ symmetry.  One can see that it
goes up and branches with the dark orange curve that corresponds to
the $(-,+)$ eigenmode at the top of the figure.  However, a careful
examination reveals that this eigenmode branches first with the
$(+,+)$ eigenmode right below it, then they split again before
branching with the light orange curve.

At a large gradient strength, nearly all plotted eigenvalues have
branched, and the associated eigenmodes are localized on one side of
the domain.  Consistently with our results, eigenvalues with positive
imaginary part correspond to eigenmodes localized on the left side of
the disk. By applying the theory of localization at a curved boundary
\cite{Moutal2019b,Swiet1994a}, one can assign to each eigenmode two
indices $(n,l)$ that describe the behavior of the eigenmode in the
directions perpendicular and parallel to the boundary.  As the order
$n$ increases, the extension of the eigenmodes along $x$ increases
until a point where they cannot be localized anymore.  This explains
why the branch points associated to larger values of $n$ occur at
larger values of $g$.

\section{Discussion}
\label{sec:discussion}

In many applications, the governing operator $\B(g)$ admits a natural
representation $\B(g) = \B(0) + g {\mathcal V}$, in which $\B(0)$ and
$g{\mathcal V}$ describe respectively an ``unperturbed'' system and
its perturbation.  For instance, in quantum mechanics, $\B(0)$ can be
the kinetic energy $-\hbar^2 \nabla^2/(2m)$ of a particle of mass $m$,
while $g{\mathcal V}$ be an applied potential.  In diffusion NMR,
$\B(0) = -\nabla^2$ and ${\mathcal V} = ix$ describe the diffusion
motion and spin precession, respectively.  As an exact computation of
the eigenvalues and eigenmodes of the operator $\B(g)$ is often
challenging, one resorts to perturbative expansions on the basis of a
simpler operator $\B(0)$.  For instance, the cumulant expansion of the
macroscopic signal is the cornerstone of the current theory of
diffusion NMR \cite{Grebenkov2007a,Kiselev2017a}.  In particular, its
first-order and second-order truncations, known as the Gaussian phase
approximation and the kurtosis model, are among the most commonly used
models for fitting and interpreting the NMR signal in medical
applications (see
\cite{Grebenkov2007a,Kiselev2017a,Jensen2005,Trampel2006,LeBihan2012,Novikov2018}
and references therein).  Despite their practical success, we argue
that the cumulant expansion and other perturbative expansions have
generally a finite radius of convergence, i.e., they unavoidably fail
to describe the quantity of interest (e.g., the macroscopic signal) at
large $g$.

In this light, the branch point with the smallest absolute value plays
a particularly important role by providing the upper bound $g_c$ on
the convergence radius $g_{\rm max}$ outside of which small-$g$
expansions fail because of the non-analyticity of the spectrum.  In
other words, such expansions diverge for any $|g| \geq g_c \geq g_{\rm
max}$.  Note that the convergence radius may be smaller than $g_c$ so
that the opposite inequality, $|g| < g_c$, does not ensure the
convergence.  A simple scaling argument yields that $g_c = \eta/L^3$,
where $L$ is a characteristic size of the domain, while $\eta$ is a
dimensionless parameter that depends on the domain (and, in general,
on its orientation with respect to the applied gradient).  Stoller
{\it et al.} determined the set of branch points of the Bloch-Torrey
operator on an interval of length $L$ with reflecting endpoints (see
Table II in \cite{Stoller1991a}).  In particular, they found
$\eta_{\rm int} = 8(\sqrt{3} \frac{27}{32} j_1^2) \approx 18.06$,
where $j_1 \approx 1.243$ is the first zero of the Bessel function
$J_{-2/3}(z)$ of the first kind%
\footnote{
Here we added the factor $8$ to the original expression for the first
branch point in \cite{Stoller1991a} to account for the fact that
Stoller {\it et al.} considered the interval of length $2$.}.
%footnote
This theoretical value is in perfect agreement with our numerical
computations shown on Fig. \ref{fig:1D}.
For a disk of diameter $L$, Fig. \ref{fig:pure_beauty} suggests $L
g_c^{1/3} \approx 3.1$, from which $\eta_{\rm disk} \approx 30$.  This
critical value implies an upper bound for the magnetic field gradient
\begin{equation}
G < G_c \approx \eta \frac{D}{\gamma L^3} \,,
\end{equation}
which is the {\it necessary} (not sufficient) condition for the
convergence of perturbative low-gradient expansions.  

Let us estimate this upper bound on the gradient for some experimental
settings.  For instance, for a diffusion NMR with hyperpolarized
xenon-129 gas in a cylinder of diameter $L$ \cite{Moutal2019b}, one
has $D \approx 3.7\cdot 10^{-5}~\mathrm{m}^2~\mathrm{s}^{-1}$ and
$\gamma \approx 7.4\cdot 10^{6}~\mathrm{T}^{-1}~\mathrm{s}^{-1}$, so
that $G_c \approx 19$~mT/m for $L = 2$~mm, and $G_c \approx 2.7$~mT/m
for $L = 3.8$~mm.  These are small gradients, which are available and
often used in most clinical MRI scanners.  As illustrated on Fig. 6
from \cite{Moutal2019b}, the lowest-order (Gaussian phase)
approximation captures correctly the macroscopic signal behavior at
$G$ below $5-10$~mT/m, and then fails.  In other words, the
lowest-order truncation of the cumulant expansion may eventually be
accurate above the upper bound $G_c$.  This is even more striking in
an experiment realized for water molecules confined between two
parallel plates \cite{Huerlimann1995a}, for which $D
\approx 2.4\cdot 10^{-9}~\mathrm{m}^2~\mathrm{s}^{-1}$, $\gamma
\approx 2.68\cdot 10^{8}~\mathrm{T}^{-1}~\mathrm{s}^{-1}$, and $L =
0.16$~mm, so that $G_c \approx 3.9\cdot 10^{-2}$~mT/m, whereas the
lowest-order expansion is applicable to up $10-15$~mT/m.  One sees
that $G_c$ tell us nothing about the applicability range of
low-order-truncated expansions.  For instance, the Gaussian phase
approximation can be accurate on a much broader range of gradients,
far beyond $G_c$.  However, any attempt to improve such a
low-order-truncated expansion by adding many next-order terms (higher
cumulants) is doomed to fail for $G > G_c$.  This is a typical feature
of an asymptotic series, which is divergent but its truncation may
yield an accurate approximation (as, e.g., in the case of the
Stirling's approximation for the Gamma function).
As branch points may exist for any bounded domain, such asymptotic
expansions should be used with caution.  The proof of existence of
branch points in general bounded domains and the numerical computation
of the convergence radius (or at least of its upper bound $g_c$)
present important perspectives for future research.

We note that the finite radius of convergence of the cumulant
expansion was earlier discussed by Fr{\o}hlich {\it et al.} for a
one-dimensional model in the limit of narrow-gradient pulses
\cite{Frohlich2006}.  In that case, the gradient pulse of duration
$t$ effectively applies a $e^{iq x}$ phase pattern across the domain
with the wavenumber $q = \gamma G t$, and the decay of the
magnetization is caused by the ``blurring'' of this pattern due to
diffusion \cite{Moutal2020b,Moutal2020d}.  In this regime, the signal
is an analytical function of $bD = q^2 D T$ (with $T$ being the
inter-pulse time \cite{Stejskal1965b}) because it is controlled by the
spectrum of the Laplace operator that does not have branch points.  As
Fr{\o}hlich {\it et al.} explained, the finite convergence radius of
the cumulant expansion is merely caused by the use of the Taylor
series of the logarithm function and is therefore related to the
smallest (in absolute value) complex value of $bD$ for which the
signal is zero.  In contrast, we argue here that the non-analyticity
of the spectrum of the Bloch-Torrey operator at branch points
intrinsically restricts the range of applicability of low-gradient
expansions in bounded domains.  Moreover, it is not the $b$-value,
which is commonly used to characterize the gradient setup, but the
gradient strength $G$ that controls the convergence radius.

\section{Conclusion}
\label{sec:conclusion}

In this paper, we investigated the spectrum of the Bloch-Torrey
operator in bounded domains and focused on its branch points at which
two eigenvalues coincide.  Using a matrix model, we explained the
origin and the square-root structure of these points.  The major
difference with respect to the Hermitian case is the non-analytical
behavior in a vicinity of a branch point, as well as a transformation
of two linearly independent eigenmodes into a single eigenmode and a
generalized eigenmode.  We argued how the geometric structure of the
eigenmodes and their symmetries drastically change below and above the
branch point, and described the consequent localization of eigenmodes
at sufficiently large gradient $g$, which is responsible for the
emergence of the localization regime in diffusion NMR
\cite{Stoller1991a,Swiet1994a,Grebenkov2018a,Moutal2019b,Moutal2020b}.
We discussed some consequences of these spectral properties, in
particular, how the branch point with the smallest amplitude
determines an upper bound on the convergence radius of perturbative
expansions such as the cumulant expansion for the macroscopic signal
in diffusion NMR experiments.

For $x$-parity symmetric domains, the existence of branch points lying
on the real axis follows from the paired structure of the eigenmodes
and eigenvalues.  However, this symmetry argument is not applicable
for arbitrary bounded domains.  To reveals the existence of branch
points in this general setting, we presented an efficient numerical
algorithm for determining the branch points in the complex plane.
This algorithm revealed typical patterns of branch points for
symmetric and asymmetric domains.  In particular, if some branch
points may lie on the real axis for symmetric domains, this is
generally not the case for asymmetric domains.  In the latter case,
even though there may be no branching for any real $g$, the
convergence radius of perturbative expansions is still finite, due to
the presence of branch points beyond the real axis.  This suggests
that a better mathematical understanding of the spectral properties of
the Bloch-Torrey operator requires considering complex-valued
gradients as well.

The focus on the Bloch-Torrey operator in bounded domains was mainly
dictated by our will to present the spectral properties in a pedagogic
way by avoiding mathematical subtleties and related technical
difficulties of more general settings.  For instance, the spectrum of
the Bloch-Torrey operator $\B(g) = -\nabla^2 - igx$ is known to be
discrete as $- igx$ is a bounded perturbation to the unbounded Laplace
operator.  However, we expect that many features presented in this
paper are rather general.  In particular, recent works brought
evidence that the spectrum of the Bloch-Torrey operator in the
exterior of a compact domain and in some periodic domains can be
discrete as well \cite{Almog2018a,Moutal2020a,Grebenkov2021a}.  This
statement may sound counter-intuitive because the spectrum of the
Laplace operator, corresponding to $g = 0$, is continuous for
unbounded or periodic domains.  In other words, the limiting behavior
of the spectral properties of $\B(g)$ in the vicinity of $g = 0$ is
singular, whereas basic perturbative approaches would fail for any
$g$.  These spectral properties remain poorly understood; in
particular, there is no information on the branch points in such
unbounded domains, except for some numerical evidence reported in
\cite{Moutal2020a}.  We believe that future investigations in this
direction will improve our understanding of the localization regime in
diffusion NMR.  Furthermore, one can study spectral branch points for
many other non-Hermitian operators by applying projectors and
resorting to finite-dimensional matrices, or by using our numerical
algorithm.

Our study brings more evidence that the current perturbative approach
to the theory of diffusion NMR should be revised.  Even though this
approach was successful for moderate gradients, it has fundamental and
practical limitations, in particular, due to the finite convergence
radius.  Such a revision may be fruitful even from a broader
scientific perspective.  In fact, the current perturbative approach
aims at eliminating the mathematical challenges by reducing the
analysis of the non-Hermitian Bloch-Torrey operator to the study of
the Hermitian Laplace operator.  Perhaps, it is time to address these
challenges in a more rigorous, non-perturbative way and to explore
much richer spectral features of the non-Hermitian Bloch-Torrey
operator to gain new theoretical insights and experimental modalities
in diffusion NMR.

\ack

The authors thank Prof. B. Helffer for fruitful discussions.
D.S.G. acknowledges the Alexander von Humboldt Foundation for support
within a Bessel Prize award.

\appendix
\section{Numerical computation of the spectrum}
\label{sec:numerics}

In this appendix, we briefly describe the numerical procedure for
computing the spectrum of the Bloch-Torrey operator.  Here we rely on
the so-called matrix formalism
\cite{Caprihan1996a,Callaghan1997a,Barzykin1998a,Sukstanskii2002a,Grebenkov2008b},
in which the differential operator $\mathcal B(g) = -\nabla^2 - igx$
is represented by an infinite-dimensional matrix $\mathsf{\Lambda} -
ig \mathsf{B}$, where $\mathsf{\Lambda}$ is the diagonal matrix formed
by the eigenvalues of the Laplace operator $-\nabla^2$ with Neumann
boundary condition, while the matrix 
\begin{equation}  \label{eq:B_def}
\mathsf{B}_{n,n'} = \int\limits_{\Omega} d\mathbf{r} \, u_n(\mathbf{r}) \, x \, u_{n'}^*(\mathbf{r})
\end{equation}
represents a linear potential $x$ in the complete basis of
eigenfunctions $u_n$ of the Laplace operator.  In other words, one can
expand an eigenmode $v_n$ of $\B(g)$ as
\begin{equation}  \label{eq:vn_un}
v_n(\r) = \sum\limits_{j} \V_{n,j}\,  u_j(\r),
\end{equation}
substitute its into the eigenvalue equation $\B(g) v_n = \lambda_n(g)
v_n$, multiply it by another Laplacian eigenfunction function
$u_k^*(\r)$ and integrate over the confining domain $\Omega$ to get
\begin{equation}
\sum\limits_j \V_{n,j} \underbrace{\int\limits_{\Omega} d\r \, \bigl(\B(g) \, u_j(\r) \bigr) \, u^*_k(\r)}_{\bigl[\mathsf{\Lambda} -
ig \mathsf{B}\bigr]_{j,k}} = \lambda_n(g) \V_{n,k} \,,
\end{equation}
where we used the orthogonality of eigenfunctions $u_k$ to simplify
the right-hand side.  In a matrix form, this set of linear equations
reads 
\begin{equation}
\V (\mathsf{\Lambda} - ig \mathsf{B}) = \mathsf{\Lambda}(g) \V, 
\end{equation}
where $\mathsf{\Lambda}(g)$ is the diagonal matrix with eigenvalues
$\lambda_n(g)$ of the Bloch-Torrey operator.  
The matrix $\mathsf{\Lambda} - ig \mathsf{B}$ is then truncated and
diagonalized numerically in a standard way.  The eigenvalues of the
truncated matrix approximate $\lambda_n(g)$, whereas the associated
eigenvectors determine the coefficients $\V_{n,j}$ needed to
reconstruct the eigenmodes $v_n$ of the Bloch-Torrey operator via
truncated expansions \eqref{eq:vn_un} over the eigenfunctions $u_n$.

The advantage of the matrix formalism is that the matrices
$\mathsf{\Lambda}$ and $\mathsf{B}$ have to be constructed only once
for a given domain $\Omega$ that facilitates computation of the
eigenvalues $\lambda_n(g)$ as functions of $g$.  For some symmetric
domains (e.g., an interval, a disk, or a sphere), the eigenvalues and
eigenfunctions of the Laplace operator are well known; moreover, the
integrals in \eqref{eq:B_def} determining the matrix $\mathsf{B}$ were
also computed exactly
\cite{Grebenkov2008a,Grebenkov2008b,Grebenkov2010a}.  We used the
explicit formulas for an interval to draw Fig. \ref{fig:1D} and those
for a disk to produce Figs.
\ref{fig:branchement_complexe_disk}(top), \ref{fig:feuilles_spectre},
and \ref{fig:pure_beauty}.  In turn, for other domains, the
eigenvalues and eigenfunctions of the Laplace operator, as well as the
integrals in \eqref{eq:B_def}, can be found numerically.  To plot
Fig. \ref{fig:branchement_complexe_disk}(bottom), we used a finite
element method implemented as a PDEtool in Matlab (see more details in
\cite{Moutal2020a,Moutal2020b}).

%\clearpage

\end{document}